\documentclass[review]{elsarticle}

\usepackage[utf8]{inputenc}
\usepackage{xcolor}
\usepackage{subfig}
\usepackage{graphicx}
\usepackage{tabularx}
\usepackage{multirow}
\usepackage{booktabs}
\usepackage{amsmath,amssymb}
\usepackage{bm}
\usepackage{makecell}
\usepackage{lineno,hyperref}

\modulolinenumbers[5]

\newcolumntype{L}[1]{>{\raggedright\let\newline\\\arraybackslash\hspace{0pt}}m{#1}}
\newcolumntype{C}[1]{>{\centering\let\newline\\\arraybackslash\hspace{0pt}}m{#1}}
\newcolumntype{R}[1]{>{\raggedleft\let\newline\\\arraybackslash\hspace{0pt}}m{#1}}

\biboptions{authoryear}

\journal{Medical Image Analysis}

\graphicspath{{figures/}}

\def\vec#1{\ensuremath{\bm{\mathit{#1}}}}

\begin{document}

\begin{frontmatter}

\title{Quantifying and Leveraging Predictive Uncertainty for Medical Image Assessment}

\author[Siemens]{Florin~C.~Ghesu\corref{cor1}}
\ead{florin.ghesu@siemens-healthineers.com}
\author[Siemens]{Bogdan~Georgescu}
\author[Siemens]{Awais~Mansoor}
\author[Siemens]{Youngjin~Yoo}
\author[Siemens]{Eli~Gibson}
\author[SiemensIndia]{R.S.~Vishwanath}
\author[SiemensIndia]{Abishek~Balachandran}
\author[Michigan]{James M. Balter}
\author[Michigan]{Yue Cao}
\author[MGH,Harvard]{Ramandeep Singh}
\author[MGH,Harvard]{Subba R. Digumarthy}
\author[MGH,Harvard]{\\Mannudeep K. Kalra}
\author[Siemens]{Sasa~Grbic}
\author[Siemens]{Dorin~Comaniciu}
\address[Siemens]{Siemens Healthineers, Digital Technology and Innovation, Princeton, NJ, USA}
\address[SiemensIndia]{Siemens Healthineers, Digital Technology and Innovation, Bangalore, India}
\address[Michigan]{University of Michigan, Department of Radiation Oncology, Ann Arbor, MI, USA}
\address[MGH]{Department of Radiology, Massachusetts General Hospital, Boston, MA, USA}
\address[Harvard]{Harvard Medical School, Boston, MA, USA}

\cortext[cor1]{Corresponding author}

\begin{abstract}

The interpretation of medical images is a challenging task, often complicated by the presence of artifacts, occlusions, limited contrast and more. Most notable is the case of chest radiography, where there is a high inter-rater variability in the detection and classification of abnormalities. This is largely due to inconclusive evidence in the data or subjective definitions of disease appearance. An additional example is the classification of anatomical views based on 2D Ultrasound images. Often, the anatomical context captured in a frame is not sufficient to recognize the underlying anatomy. Current machine learning solutions for these problems are typically limited to providing probabilistic predictions, relying on the capacity of underlying models to adapt to limited information and the high degree of label noise. In practice, however, this leads to overconfident systems with poor generalization on unseen data. To account for this, we propose a system that learns not only the probabilistic estimate for classification, but also an explicit uncertainty measure which captures the confidence of the system in the predicted output. We argue that this approach is essential to account for the inherent ambiguity characteristic of medical images from different radiologic exams including computed radiography, ultrasonography and magnetic resonance imaging. In our experiments we demonstrate that sample rejection based on the predicted uncertainty can significantly improve the ROC-AUC for various tasks, e.g., by 8\% to 0.91 with an expected rejection rate of under 25\% for the classification of different abnormalities in chest radiographs. In addition, we show that using uncertainty-driven bootstrapping to filter the training data, one can achieve a significant increase in robustness and accuracy. Finally, we present a multi-reader study showing that the predictive uncertainty is indicative of reader errors.
\end{abstract}


\end{frontmatter}

\section{Introduction}

The interpretation of medical images is an essential task in the practice of a radiologist, yet challenging due to inconclusive and ambiguous image information, image artifacts, occlusions and more. Consider the example of chest radiography, a key imaging exam that enables the early detection of abnormalities in the lungs, heart or chest wall~\citep{Rajpurkar2018}. Driven by the need to improve and accelerate the interpretation of such images, several deep learning solutions have been proposed for the automatic classification of radiographic findings~\citep{Wang2017,Guendel2018,Guendel2019,Yao2018}. However, development, training and validation of these solutions are challenged by significant inter-rater variations in detection and classification of such radiographic findings~\citep{Rajpurkar2018}. This can be caused by inconclusive evidence in the data or subjective definitions of the appearance of different findings. Similar challenges are faces in other applications as well, e.g., the view-classification of abdominal ultrasound images or assessment of brain metastases in magnetic resonance (MR) scans of the brain. We argue that modeling this variability when designing a system for assessing this type of data is essential -- an aspect which was not considered in previous work.\smallskip

Using principles of information theory and subjective logic~\citep{Josang2016} based on the Dempster-Shafer framework for modeling of evidence~\citep{Dempster1968}, we present a method for training a parametric model that generates both an image-level class probability and a corresponding uncertainty measure. We evaluate this method on the image-level labeling of abnormalities on chest radiographs, the view-classification of abdominal ultrasound images, and detection of small brain metastases in brain MR scans. There, we demonstrate that one can effectively use the uncertainty measure to avoid returning a prediction on cases with highest uncertainty, thereby consistently achieving a more accurate classification on the remaining cases. Also, we propose uncertainty-driven bootstrapping as a means to filter training samples with highest predictive uncertainty in order to improve robustness and accuracy on unseen data. Finally, we empirically show that the uncertainty measure can distinguish radiographs with correct and incorrect labels according to a multi-radiologist-consensus study. This correlation indicates the potential of the uncertainty metric to help build trust between the user and the system. This paper is an extended version of our work presented in \citep{Ghesu2019}.

\section{Background and Motivation}

\subsection{Machine Learning for Abnormality Assessment}

\textbf{Assessment of Chest Radiographs}: The open access to the ChestX-Ray8 dataset~\citep{Wang2017} of chest radiographs has led to a series of publications that propose machine learning based systems for abnormality detection and classification. With this dataset, \cite{Wang2017} evaluated several state-of-the-art convolutional neural network architectures to address this problem and reported a baseline average area under receiver operating characteristic curve (ROC-AUC) of 0.75. This performance level was increased by \cite{Islam2017} using an ensemble of classification models. Further improvements were achieved by modeling the correlation between different abnormalities based on prevalence and co-morbidity~\citep{Yao2017, Yao2018}, and explicitly integrating information observed in lateral/oblique radiographs paired with the frontal projection radiographs~\citep{Rubin2018}. An alternative method focused on driving the attention of the learning model on the image sub-regions that are most relevant for the considered abnormalities~\citep{Guan2018}. \cite{Cai2018} proposed an attention mining strategy to identify regions with abnormalities and showed that it significantly outperforms heuristic approaches, such as class activation maps. A curriculum learning method based on quantified disease severity-levels was proposed as alternative~\citep{Tang2018}.

State-of-the-art results on the official split of the ChestX-Ray8 dataset are reported in~\cite{Guendel2018}, including the follow-up work presented in~\cite{Guendel2019}. Using multi-task learning coupled with a location-aware dense neural network learning architecture, an average ROC-AUC of 0.81 was achieved. On the official split of the dataset from the Prostate, Lung, Colorectal and Ovarian (PLCO) Cancer Screening Trial~\citep{PLCO}, an average performance of 0.88 (ROC-AUC) was reported for 12 different abnormalities.\smallskip

In light of all these publications, a recent study compared the performance of such an AI system with the performance of 9 practicing radiologists~\citep{Rajpurkar2018}. The authors selected 6 board-certified radiologists with an average experience of 12 years (ranging between 4 and 28 years) and 3 senior radiology residents. The 9 readers originated from 3 academic institutions. While the study indicates that the system can surpass human performance, it also highlights the high variability among different expert radiologists for the interpretation of chest radiographs. The reported average specificity of the readers was very high (over 95\%), while their average sensitivity was 50\%$\,\pm$\,8\%. Such a large inter-rater variability leads to several questions: How can one obtain accurate ground truth data? To what extent does the label noise affect the training process and the system performance? The mentioned reference solutions do not consider this variability. In practice, this strategy typically leads to models with overconfident predictions and limited generalization on unseen data. In this context, \cite{Irvin2019} recently presented a new public dataset with image level annotations and additional uncertainty labels that can be taken into account during system training.\smallskip

\textbf{Detection of Brain Metastases}: An additional example is the localization of brain metastases for treatment selection, planning and longitudinal monitoring. Contrast-enhanced magnetization-prepared rapid acquisition with gradient echo (MPRAGE) scans enable the detection and segmentation of brain metastases, which serves as essential information in guiding radiosurgery protocols and other treatment decisions. However, identifying and segmenting small metastases (i.e., $\le 1 \text{cm}^3$) is challenging and there is a significant degree of ambiguity in assessing and manually annotating such small metastases~\citep{pope2018brain}. Computer-aided assistance could have a significant impact in the staging, selection and implementation of treatment, and assessment of therapeutic response; however, the ambiguity in the visual appearance and hence in the reference data annotation also limit the performance of machine learning systems trained for this task.

\subsection{Machine Learning for Image-View Classification}

Abdominal ultrasonography (US) is a commonly performed imaging test for a variety of ailments.  High patient throughput and decreasing reimbursement can lead to errors or lack of anatomic markup and orientation in the acquired images~\citep{vannetti2015usability}. A typical abdominal exam comprises of a trained sonographer navigating to and capturing a series of views of abdominal organs, freezing the view, and recording clinically relevant measurements. The classification of these views depending on the underlying anatomy is a challenging problem; recent studies indicated a inter-rater agreement of less than 80\%. This is mainly because of the quality of the image or symmetry confusion (e.g., between left and right kidney). The substantial manual interaction is not only burdensome for the operator and substantially lowers the workflow efficiency, but also introduce user bias to the acquired patient data which has led to a series of publications on how to optimize US-based workflow screening~ \citep{xu2018less,lin2019multi, otey2006automatic, aschkenasy2006unsupervised}. This uncertainty poses a challenge to a machine learning system in solving the task.

\subsection{Principles of Uncertainty Estimation} 

Explicitly quantifying the classification uncertainty based on observed data is a principled strategy to address the aforementioned challenges. Early contributions in this field were based on Bayesian estimation theory to measure model uncertainty~\citep{Hinton1993, Mackay1992}. In the context of deep learning, techniques such as variational dropout~\citep{Molchanov2017, Gal2016, Kingma2015} have been proposed to approximate Bayesian learning, while better coping with the high computational requirements for large hierarchical models. Benefits are demonstrated in~\citep{Kuo2019}. Alternatively, ensembles of deep learning models have been proposed by~\cite{Laks2017} to implicitly model both epistemic and aleatoric uncertainty (see Figure~\ref{fig:general}). Nonetheless, the computational complexity of these methods remains suboptimal for training large models that are often used in practice.

\section{Proposed Method}

We propose a model for joint sample classification and predictive uncertainty estimation, following the Dempster-Shafer theory of evidence~\citep{Dempster1968} and principles of subjective logic~\citep{Josang2016}. This research is inspired by the work of~\cite{Sensoy2018}. 

\subsection{Modeling the Predictive Uncertainty}

We focus on the problem of binary classification, and define the per-class estimated probabilities of an arbitrary data sample $x \in \mathcal{R}^D$ as $\hat{p}_+(x)$ (for the positive class) and $\hat{p}_-(x)$ (for the negative class), and estimated predictive uncertainty as $\hat{u}(x)$, with $p_+, p_-, u : \mathcal{R}^D \rightarrow [0,1], D > 0$.

\begin{figure}[t]
\includegraphics[width=12cm]{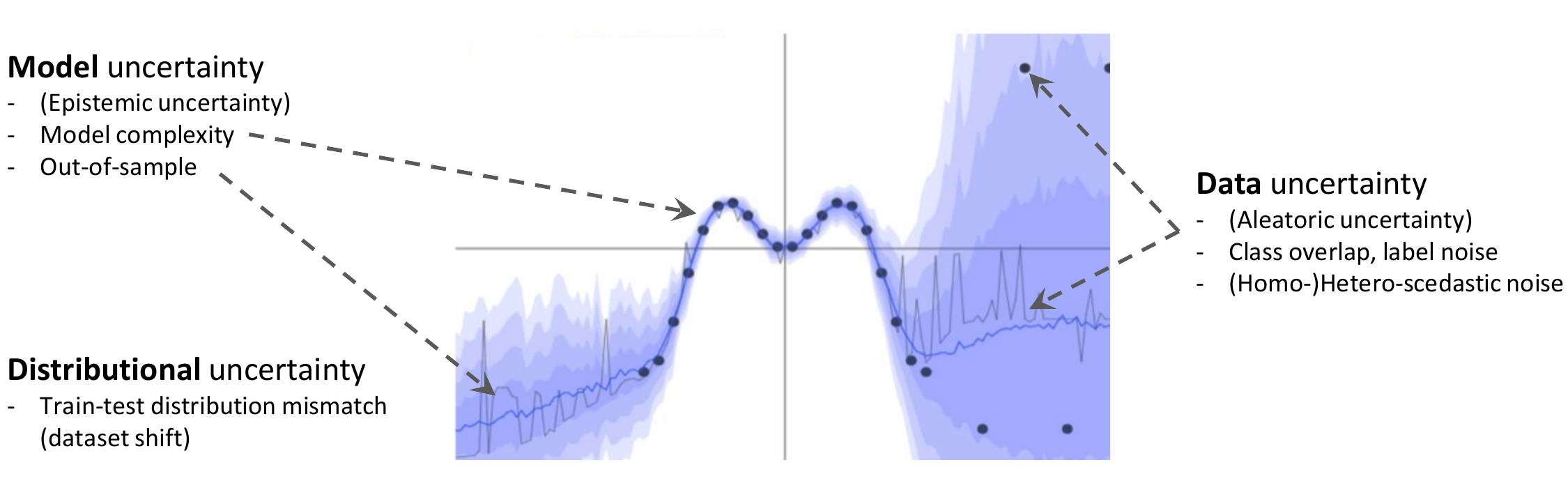}
\caption{Schematic visualization of different types of uncertainty in a functional regression example. Data uncertainty refers to inherent class overlap or label noise, a property of the underlying problem; while model uncertainty, also called epistemic uncertainty, describes the model complexity and out-of-sample behavior. Epistemic uncertainty is reducible as the size of the data is increased. Finally, distributional uncertainty is the result of a systematic difference between the distribution of the training data and the distribution of the test data, also called dataset shift~\citep{Quionero2009}. Function plot is generated using the script provided by~\cite{PLOT}.\label{fig:general}}
\end{figure}

\begin{figure*}[t]
\centering
\subfloat[Confident negative prediction]{
\includegraphics[height=4cm]{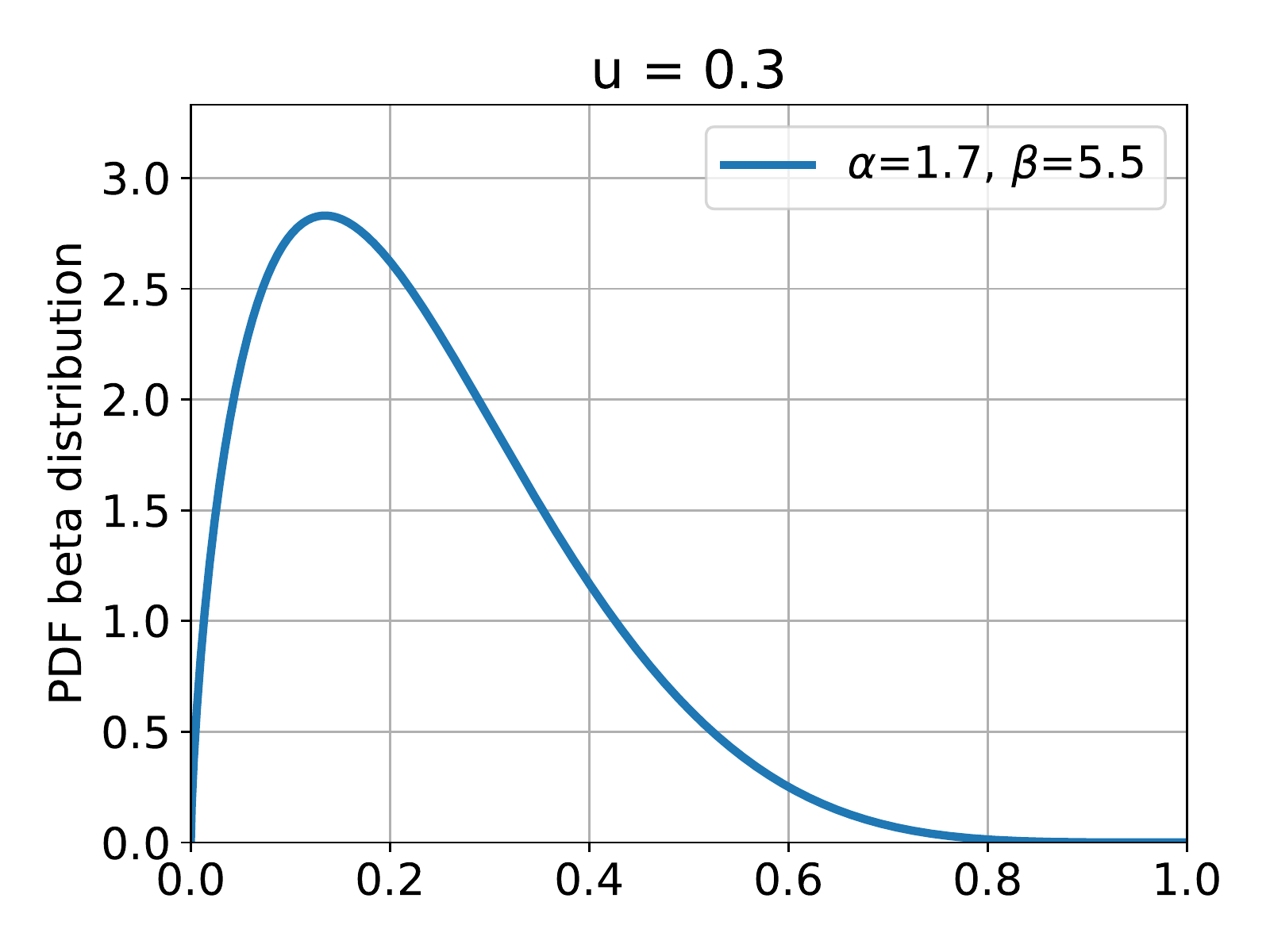}
}\hspace{-.5cm}
\subfloat[Confident positive prediction]{
\includegraphics[height=4cm]{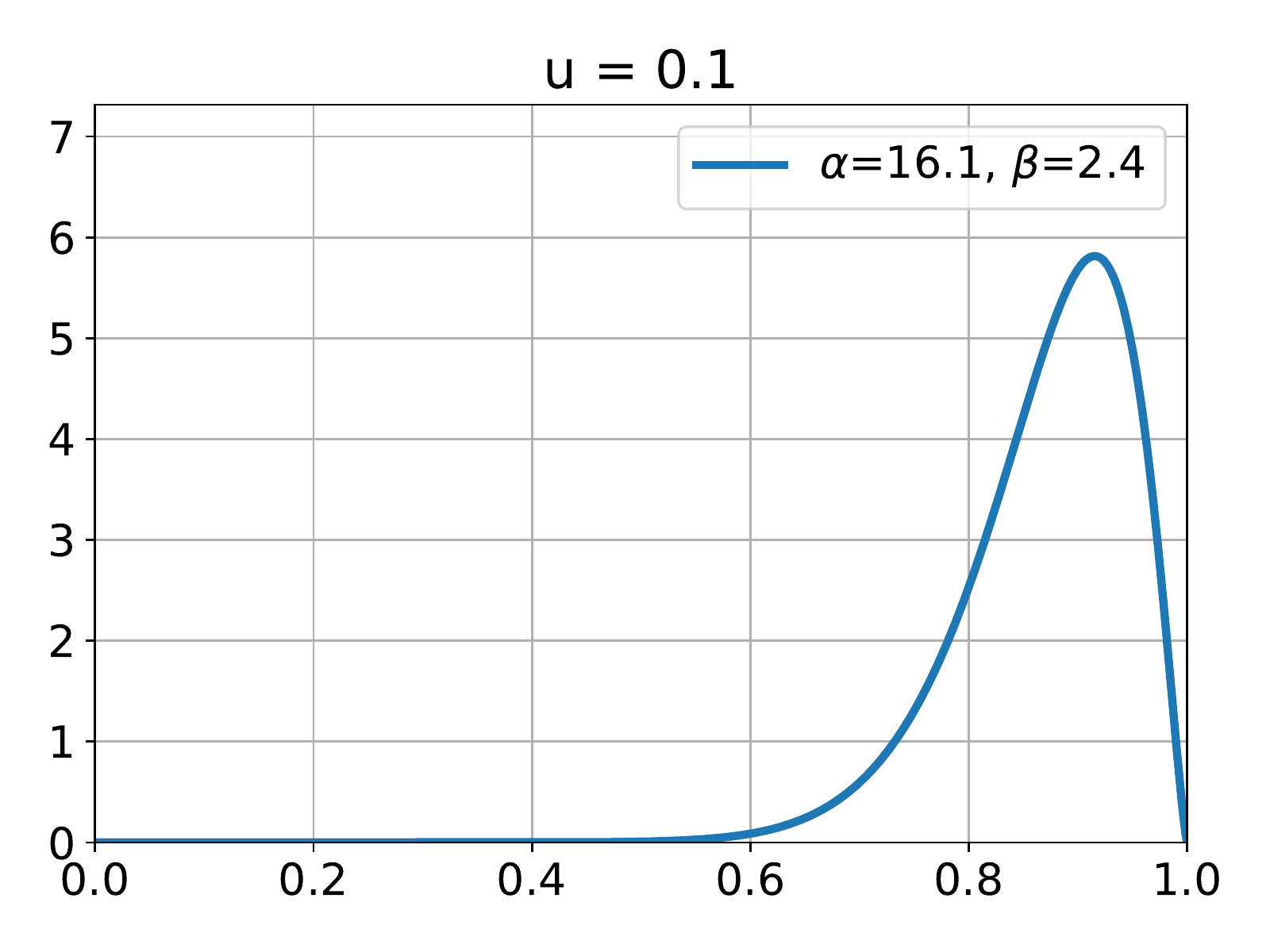}
}\\
\subfloat[High uncertainty prediction]{
\includegraphics[height=4cm]{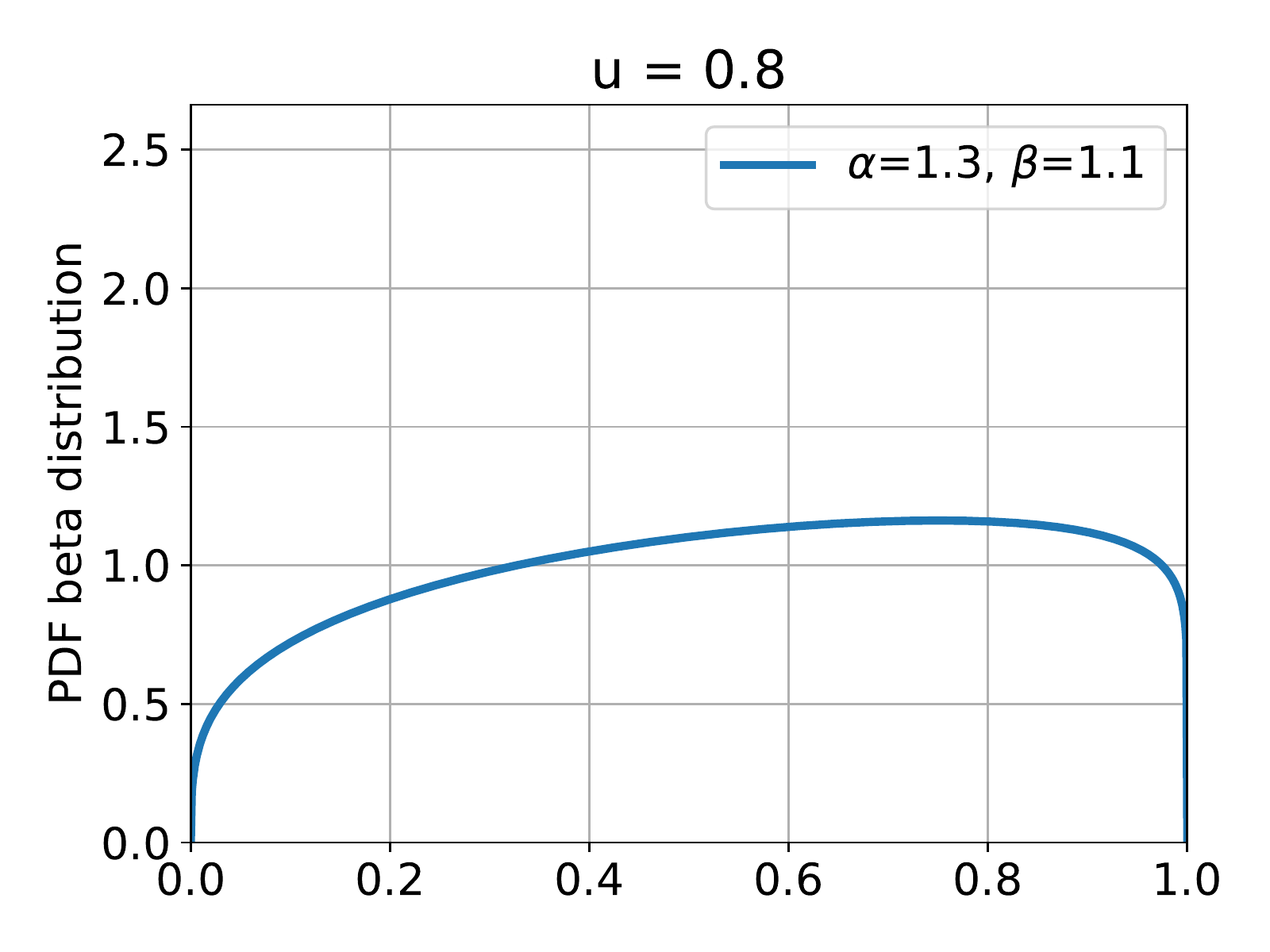}
}\hspace{-.5cm}
\subfloat[Confident 50\% prediction]{
\includegraphics[height=4cm]{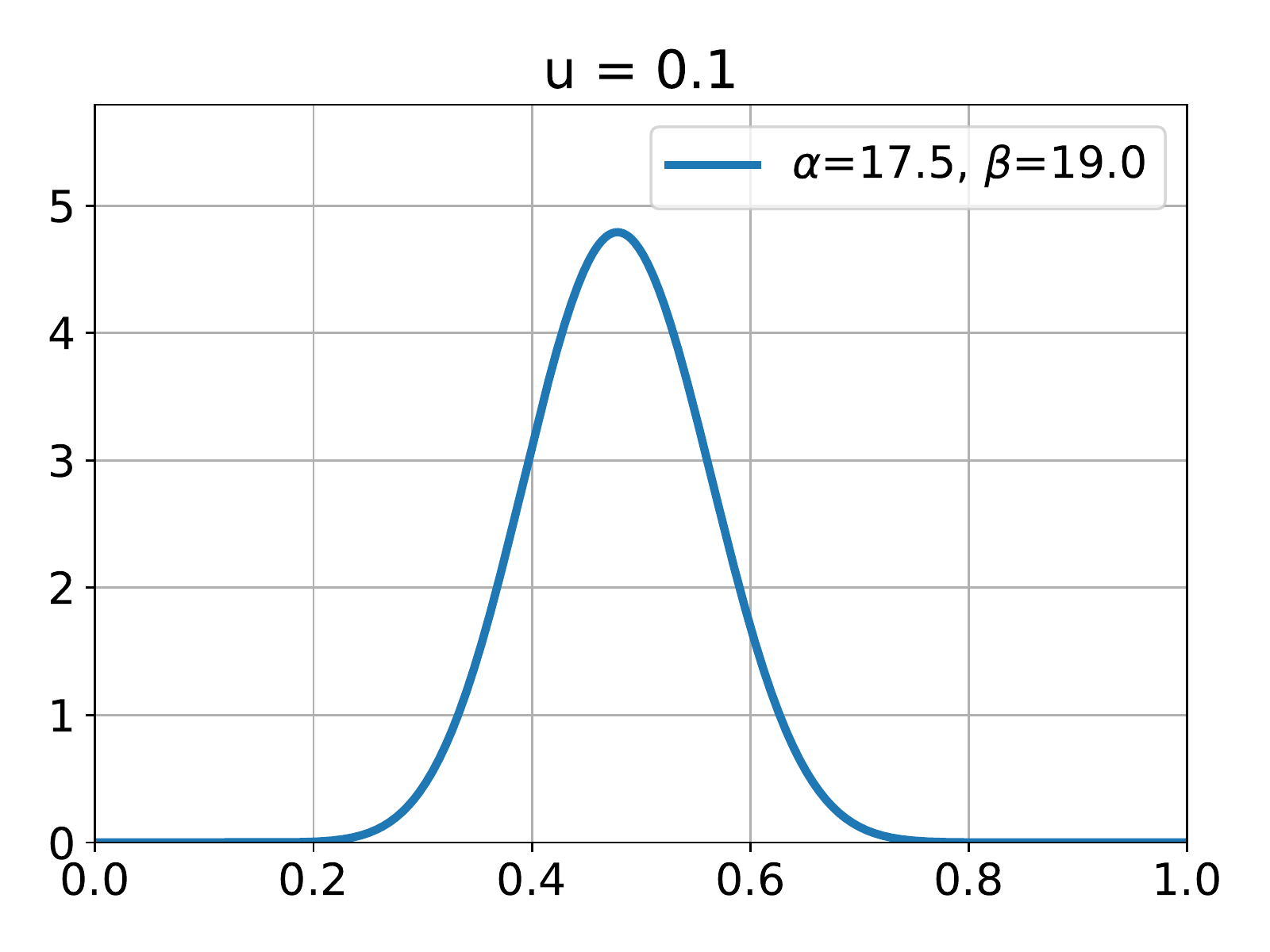}
}
\qquad
\caption{Probability density function of the beta distribution: example parameters ($\alpha, \beta$) modeling confident and uncertain predictions. Note in case (d) that the model can predict with high confidence and approximately random chance probability. This can occur in applications where there is inherent class overlap.\label{fig:beta}}
\end{figure*}

The classification problem is reduced to a problem of estimating belief values (also called belief masses) that indicate the membership of a sample $x$ to a specific class~\citep{Dempster1968,Josang2016}. We denote $b^+(x)$ and $b^-(x)$ the belief values for the positive and negative class, respectively. In this theoretical framework, these values are computed from evidence values $e^+(x), e^-(x) \ge 0$, that indicate based on features of $x$ the likelihood of it being classified in the positive, or the negative class:
\begin{equation}
\begin{aligned}
    b^+(x) &= \frac{e^+(x)}{E(x)}\\
    b^-(x) &= \frac{e^-(x)}{E(x)},\\
    \text{where\,\,} E(x)&=e^+(x)+e^-(x) + 2,
\end{aligned}
\end{equation}
$E(x)$ denoting the total evidence. With these variables, one can also quantify the so called uncertainty mass, which is defined as: $u(x) = 1 - b^+(x) - b^-(x)$. For the considered binary classification setting, we propose to model the distribution of such evidence values using the beta distribution, defined by two parameters $\alpha$ and $\beta$ as:
\begin{equation}
    f(y;\alpha,\beta) = \frac{\Gamma(\alpha+\beta)}{\Gamma(\alpha)\Gamma(\beta)}y^{\alpha - 1}(1 - y)^{\beta - 1},
\end{equation}
where $\Gamma$ denotes the gamma function and $\alpha, \beta > 1$, with $\alpha = e^+(x) + 1$ and $\beta = e^-(x) + 1$. The per-class probabilities can be derived as:
\begin{equation}
\begin{split}
p^+(x) &= \alpha/E(x)\\
p^-(x) &= \beta/E(x)
\end{split}
\end{equation} 
Figure~\ref{fig:beta} visualizes the beta distribution for different values of $\alpha$ and $\beta$. This model implicitly measures both epistemic and aleatoric uncertainty. More details about different types of uncertainty and their properties are depicted in Figure~\ref{fig:general}.

\subsection{Learning to Predict Uncertainty from Labeled Data}

Let us assume a labeled training dataset is given as $\mathcal{D} = \{\vec{I}_k,y_k\}_{k=1}^{N}$, consisting of $N$ pairs of images $\vec{I}_k$ with a binary class assignment $y_k\in\{0,1\}$. We propose to use a parametric model -- a deep convolutional neural network to estimate the per-class evidence values from the image data. Let $\vec{\theta}$ denote the parameters of this model. The evidence values are estimated as: $[\hat{e}^+_k, \hat{e}^-_k] = \mathcal{R}(\vec{I}_k;\vec{\theta})$, where $\hat{e}^+_k, \hat{e}^-_k$ denote the estimated evidence values for the positive/negative class for sample $\vec{I}_k$ and $\mathcal{R}$ denotes the model as a functional component.

Using maximum likelihood estimation, one can learn the network parameters $\hat{\vec{\theta}}$ by optimizing the Bayes risk of the class predictor $p_k$ with a beta prior distribution:
\begin{equation}
    \mathcal{L}^{data}_k = \int \|\vec{y}_k - \vec{p}_k\|^2 \frac{\Gamma(\alpha+\beta)}{\Gamma(\alpha)\Gamma(\beta)}p_k^{\alpha - 1}(1 - p_k)^{\beta - 1} d\vec{p}_k,
\label{eq:lossmse}
\end{equation}
where $\Gamma$ denotes the gamma function, $p_k = p_k^+$, $k\in\{1,\ldots,N\}$ denotes the index of the training example from dataset~$\mathcal{D}$; and $\vec{p}_k, \vec{y}_k$ represent the predicted probability and label in vector form for the training sample $k$ (label is a one-hot encoding for the two considered classes, while the prediction is defined as $\vec{p}_k=[p^+(x), p^-(x)]^\top$). $\mathcal{L}^{data}_k$ defines the goodness of fit. Using linearity properties of the expectation, Eq.~\ref{eq:lossmse} becomes:
\begin{equation}
    \mathcal{L}^{data}_k = (y_k - \hat{p}_k^{\,+})^2 + (1 - y_k - \hat{p}_k^{\,-})^2 + \frac{\hat{p}_k^{\,+}(1 - \hat{p}_k^{\,+}) + \hat{p}_k^{\,-}(1 - \hat{p}_k^{\,-})}{E_k + 1},
\end{equation}
where $\hat{p}_k^{\,+}$ and $\hat{p}_k^{\,-}$ denote the probabilistic prediction. In this equation, the first two terms measure the goodness of fit, and the last term encodes the variance of the prediction~\citep{Sensoy2018}.

A regularization term is added to the loss function to penalize low uncertainty predictions for data samples with limited/low per-class evidence values. We denote this term as $\mathcal{L}^{reg}_k$ and define as the relative entropy, i.e., the Kullback-Leibler divergence, between the beta distributed prior term and the beta distribution with total uncertainty ($\alpha, \beta = 1$). In this way, we account for cost deviations from the total uncertainty state (i.e., $u = 1$), which do not contribute to the data fit~\citep{Sensoy2018}:
\begin{equation}
\begin{split}
    \mathcal{L}^{reg}_k &= \text{KL}\left (f(\hat{p}_k;\tilde{\alpha}_k,\tilde{\beta}_k)\|f(\hat{p}_k;\langle 1,1\rangle)\right)\\
    &=\mathbb{E}_{f(\hat{p}_k;\tilde{\alpha}_k,\tilde{\beta}_k)} \left(\log f(\hat{p}_k;\tilde{\alpha}_k,\tilde{\beta}_k) - \log f(\hat{p}_k;\langle 1,1\rangle)\right),
\end{split}
\end{equation}
where $\hat{p}_k = \hat{p}_k^{\,+}$, with $(\tilde{\alpha}_k, \tilde{\beta}_k)=(1, \beta_k)$ for $y_k = 0$ and $(\tilde{\alpha}_k, \tilde{\beta}_k)=(\alpha_k, 1)$ for $y_k = 1$. Removing additive constants and using properties of the logarithm function, the regularization term becomes:
\begin{equation}
    \mathcal{L}^{reg}_k \approx \log\frac{\Gamma(\tilde{\alpha}_k+\tilde{\beta}_k)}{\Gamma(\tilde{\alpha}_k)\Gamma(\tilde{\beta_k})} + \sum_{x\in\{\tilde{\alpha}_k, \tilde{\beta}_k\}} (x - 1)\left(\psi(x) - \psi(\tilde{\alpha}_k + \tilde{\beta}_k)\right),
\end{equation}
where $\psi$ denotes the digamma function and $k\in\{1,\ldots,N\}$ is the sample index. In this context, we define the total loss: 
\begin{equation}
\mathcal{L} = \sum_{k=1}^{N}\mathcal{L}^{data}_k + \lambda\,\mathcal{L}^{reg}_k.
\label{eq:loss}
\end{equation}

Using the stochastic gradient descent method, the total loss $\mathcal{L}$ is optimized on the training set~$\mathcal{D}$ with the annealing coefficient $\lambda\in\mathbb{R}$ starting at a small value (i.e., $\lambda_0 = 0.1$) and gradually increased during training.\smallskip

An adequate sampling of the underlying distribution is essential to ensure stability during training and ensure a robust estimation of evidence value. We empirically found dropout~\citep{Srivastava2014} to be a simple and very effective strategy to address this problem. Through the random deactivation of neurons, dropout emulates an ensembles of deep models enabling an effective sampling during training. In contrast, an explicit ensemble of $M$ independently trained models $\{\hat{\vec{\theta}}_k\}_{k=1}^{M}$ may be used. Following the principles of deep ensembles~\citep{Laks2017}, the per-class evidence can be computed from the ensemble estimates $\{e^{(k)}\}_{k=1}^{M}$ via averaging. In our experiments we empirically found that there is no significant difference between the two approaches. More details can be found in Section~\ref{sec:experiments}.

\subsection{Uncertainty-driven Bootstrapping} 

Given a dataset $\mathcal{D} = \{\vec{I}_k,y_k\}_{k=1}^{N}$, let us assume $\hat{\vec{\theta}}$ denotes the estimated model parameters which are used to measure the predictive uncertainty for each sample $k$ as $\hat{u}(\vec{I}_k)$. An efficient strategy to filter the dataset $\mathcal{D}$ with the target of reducing label noise is to eliminate a fraction $\epsilon \in\left(0,1\right)$ of samples with highest uncertainty. Without loss of generality, let us reorder the samples in $\mathcal{D}$ in descending order according to the predictive uncertainty value as follows: $\hat{u}(\vec{I}_1) \ge \hat{u}(\vec{I}_2) \ge \ldots \ge \hat{u}(\vec{I}_N)$. We define the selected subset as:
\begin{equation}
    \mathcal{D}_{\epsilon} = \{\vec{I}_k,y_k\}\,\,\text{where}\,\, k\in\{1,\ldots,\lfloor(1-\epsilon)\cdot N\rfloor\}.
\end{equation}

The hypothesis is that by retraining the model on dataset $\mathcal{D}_\epsilon$ one can increase the robustness during training and improve its performance on unseen data. Please note, the fraction of eliminated samples, i.e., the value $\epsilon$, is highly dependent on the prior probability of label noise, the problem complexity and the capacity of the learning model to capture the underlying distribution of the data.

\subsubsection{Relation to Robust M-Estimators} Conceptually, one can reformulate this strategy to optimizing the cost function defined in Equation~\ref{eq:loss} using a per-sample multiplicative weight $w_k(\vec{I}_k)$ determined based on the estimated uncertainty $\hat{u}(\vec{I}_k)$. As such, this weight is proportional to the so called inlier noise with the per-sample loss $\mathcal{L}_k$ bounded to $[0,1]$. This can be regarded as a robust M-estimator described in more detail in~\citep{Meer2004}. As this is part of our ongoing work, we do not include experiments related to this approach in this paper.

\section{Experiments and Results}
\label{sec:experiments}

We investigated the performance and properties of our method on three different problems, the classification of abnormalities in frontal chest radiographs, the view-classification of abdominal ultrasound images and the detection of small metastases in MR scans of the brain. 

\subsection{Assessment of Chest Radiographs}

We considered several radiographic findings and abnormalities including calcified nodules (which are often granulomas), fibrosis, scaring, osseous lesions, cardiac abnormalities (e.g., enlarged cardiac silhouette which can suggest cardiomegaly) and pleural effusions (accumulation of fluid in the pleural space). Often these abnormalities co-occur.

\subsubsection{Dataset and Setup} 

We used two public datasets, the ChestX-Ray8 dataset published by~\cite{Wang2017} and the dataset from the Prostate, Lung, Colorectal and Ovarian (PLCO) Cancer Screening Trial~\citep{PLCO}. They contain a series of frontal chest radiographs in anterior-posterior (AP) or posterior-anterior (PA) view. Each image is associated with a binary labels indicating the presence of the considered abnormalities. The ChestX-Ray8 dataset contains 112,120 images from over 30,000 patients, with binary label annotations for 14 findings. These were automatically generated by parsing radiological reports using natural language processing (NLP) software~\citep{Wang2017}. On the other hand, the PLCO dataset was built as part of a screening trial, containing 185,421 images from over 55,000 patients and covering 12 different abnormalities. More details are provided in Table~\ref{tab:CXRdata}.

\begin{table}
\centering
\caption{Assessment of chest radiographs: dataset statistics}
\begin{tabular}{C{3.8cm} C{3.5cm} C{3.5cm}} 
& \makecell{ChestX-Ray8\\ \citep{Wang2017}} & \makecell{PLCO\\ \citep{PLCO}}\\
\hline
Number of images & 112,120 &  185,421 \\
Number of patients &  30,805 & 56,071 \\
Avg. images per patient & 3.6 & 3.3 \\
Image size & $1024\times1024$ & $\sim2500\times2100$ \\
\bottomrule
\label{tab:CXRdata}
\end{tabular}
\end{table}

We selected location-aware dense networks~\citep{Guendel2018, Guendel2019} as reference method. On the official split of the ChestX-Ray8 dataset this method achieves an average ROC-AUC of $0.81$ (for comparison, related competing methods report average scores of $0.75$~\citep{Wang2017} and $0.77$~\citep{Yao2018}. On the official split of the PLCO dataset the reported average performance is higher, at a ROC-AUC of $0.88$. We hypothesize that this difference in performance is explained by the better quality of the labels in the PLCO dataset. More details on this aspect are provided through this section. We also investigated the benefits of using deep ensembles instead of dropout to improve the sampling ($M=5$ models were trained on random subsets of 80\% of the training data; we refer to this method with the keyword [ens]).

Random subsets of images were selected from both datasets to be used for testing. These images were interpreted and manually labeled by multiple radiologists. For the PLCO dataset, $565$ testing chest radiographs were selected and annotated by $2$ board-certified expert radiologists. The final label for each image was determined using a majority vote among $3$ opinions, i.e., the $2$ reads from the aforementioned radiologists and the original label of the image - established during the cancer screening trial. For the ChestX-Ray8 dataset, 689 images were selected for testing and read by 4 board-certified radiologists. For each image, the label was decided by a consensus discussion. For both datasets, the remaining data was split at patient level in 90\% training and 10\% validation. All images were rescaled to $256\times 256$ using bilinear interpolation.\smallskip

\subsubsection{Model Architecture and Training} We used a DenseNet-121 architecture~\citep{Huang2017} and inserted a dropout layer ($\text{rate}=0.5$) after the last convolutional layer. A fully connected layer with ReLU activation units completes the mapping between the input image $\vec{I}$ and the outputs $\alpha$ and $\beta$. A systematic grid search was used to find the optimal configuration of training meta-parameters: learning rate of $10^{-4}$, around 12 training epochs -- using an early stop strategy with a patience of 3 epochs and a batch size of $128$.\smallskip

\subsubsection{Uncertainty-driven Sample Rejection}

Given predictive uncertainty $\hat{u}(\vec{I})$ estimates for each sample $\vec{I}$ of a testing set, we propose to use this measure for sample rejection, i.e., set a threshold $u_t$ and configure the system to not output its prediction on any cases with an expected uncertainty larger than $u_t$. One can view this as a system that is empowered to answer: \textit{"I don't know for sure"}. Recall, the predictive uncertainty is an additional measure to the class probability, with increased values on out-of-distribution cases under the given model. Formally, we refer to the degree of sample rejection using the term \textbf{coverage}, as an expected percentile of cases to be rejected. For example, at a coverage of 100\%, the system outputs its prediction on all cases, while at a coverage of 80\% the system avoids outputting its prediction on 20\% of the cases with highest uncertainty. 

\begin{figure}[t]
\centering
\includegraphics[height=3.7cm]{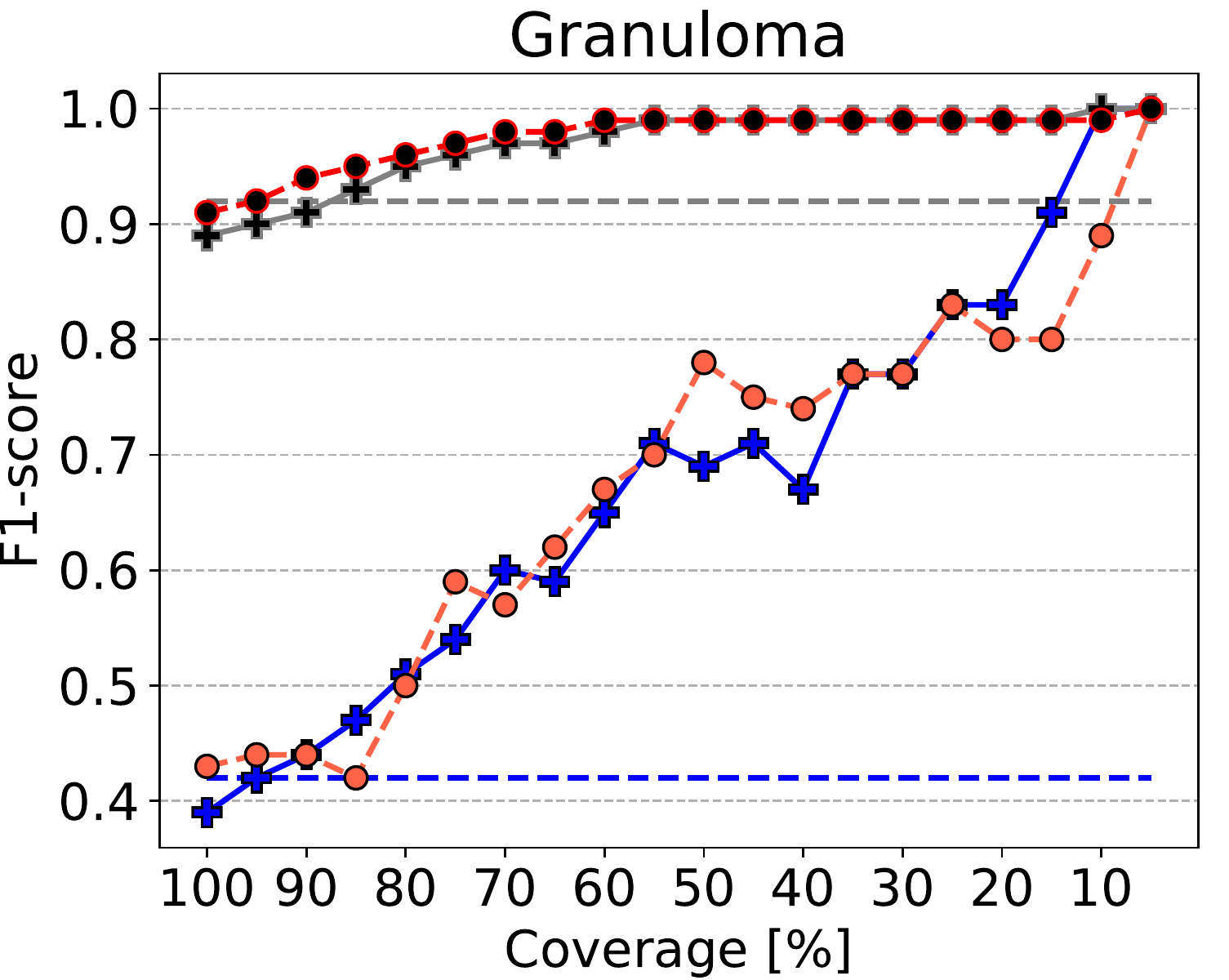}
\includegraphics[height=3.7cm]{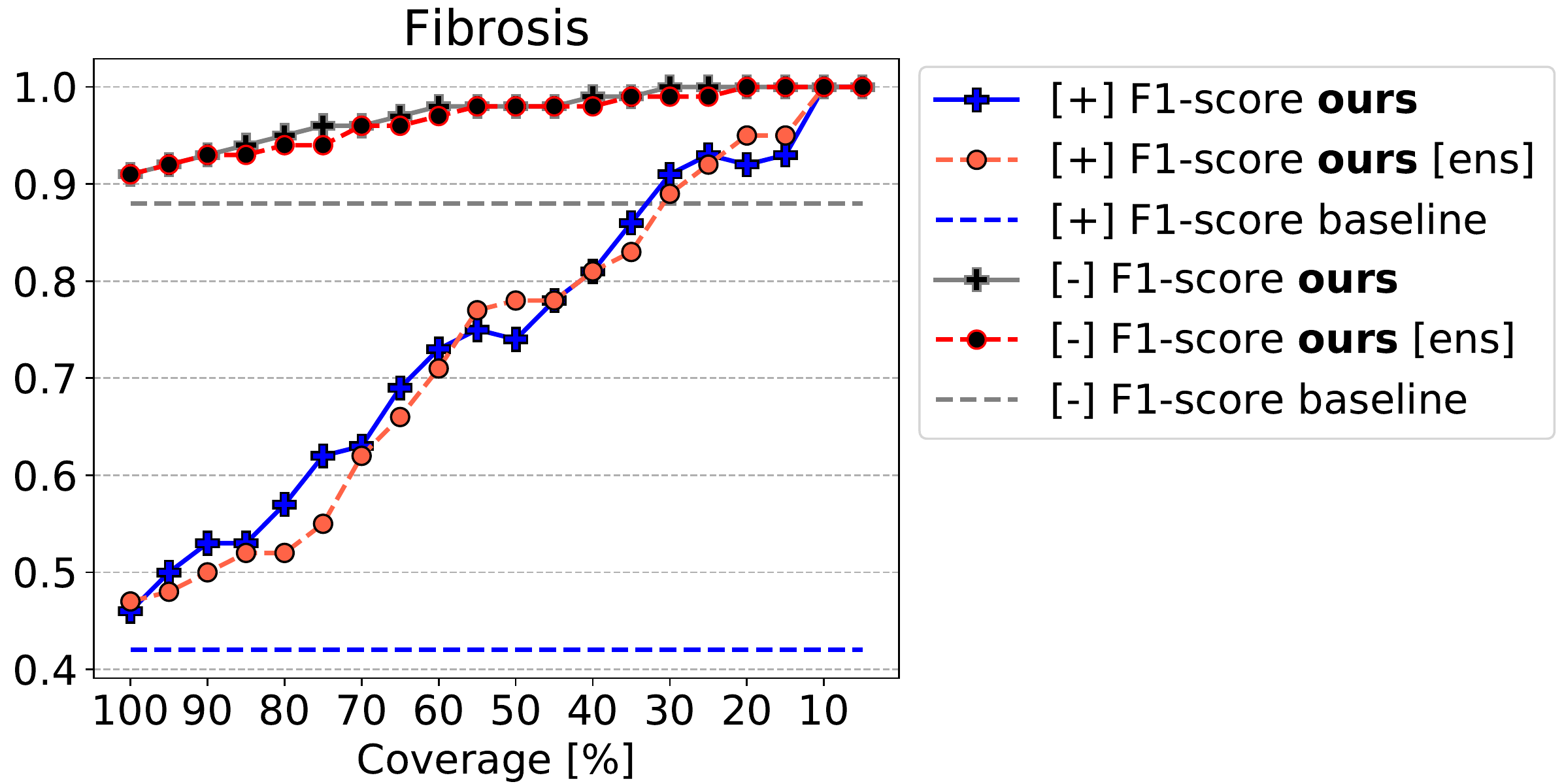}
\caption{For granuloma and fibrosis based on the PLCO~\citep{PLCO} testing set: F1-scores for the positive (+) and negative (--) classes as a function of the sample coverage, determined using the estimated uncertainty. The horizontal dashed lines denote the reference performance of the baseline method~\citep{Guendel2018}. The working point of the reference method is chosen such that the average of the per-class F1 scores is maximized. For our model, the decision threshold was set at 0.5.\label{fig:samplereject}}
\end{figure}

With this strategy one can significantly increase in system accuracy compared to the state-of-the-art on the remaining cases, as reported in Table~\ref{tab:results} and Figure~\ref{fig:samplereject}. For example, for the identification of calcified nodule or granuloma, a rejection rate of 25\% leads to an increase of over 20\% in the micro-average F1 score. We found no significant difference in average performance when using ensembles (see Figure~\ref{fig:samplereject}).

\begin{table}[t]
\centering
\caption{Our method calibrated at sample coverage rates of 100\%, 90\%, 75\% and 50\%. Testing is performed on the PLCO dataset~\citep{PLCO}. Lesion refers to osseous lesions.\label{tab:results}}
\begin{tabular}{L{1.9cm} C{2cm} C{1.4cm} C{1.4cm} C{1.4cm} C{1.4cm}}
&\multicolumn{5}{c}{\textbf{ROC-AUC}}\\
\cmidrule{2-6}
\textbf{Finding}&\cite{Guendel2018}&\textbf{Ours} [100\%]&\textbf{Ours} [90\%]&\textbf{Ours} [75\%]&\textbf{Ours} [50\%]\\
\midrule
Granuloma&0.83&0.85&0.87&\textbf{0.90}&\textbf{0.92}\\
Fibrosis&0.87&0.88&0.90&\textbf{0.92}&\textbf{0.94}\\
Scaring&0.82&0.81&0.84&\textbf{0.89}&\textbf{0.93}\\
Lesion&0.82&0.83&0.86&\textbf{0.88}&\textbf{0.90}\\
Cardiac Ab.&0.93&0.94&0.95&\textbf{0.96}&\textbf{0.97}\\
\midrule\midrule
\textbf{Average}&0.85&0.86&0.89&\textbf{0.91}&\textbf{0.93}\\
\bottomrule
\end{tabular}
\end{table}

\begin{figure}[t]
\centering
\includegraphics[width=10cm]{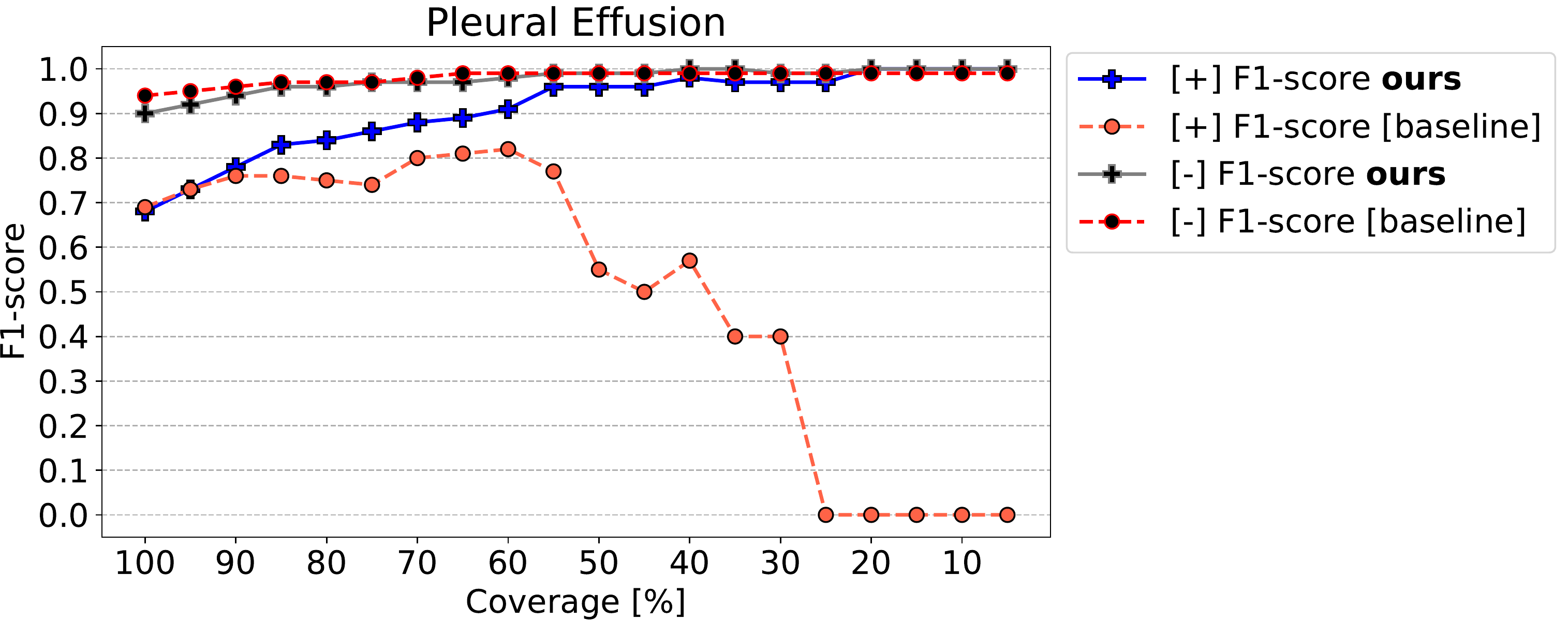}
\caption{F1-scores for the positive (+) and negative (--) classes as a function of the sample coverage, on the example of pleural effusion on the ChestX-Ray8~\citep{Wang2017} testing set. The baseline performance is determined using the method of~\cite{Guendel2018}. Please note: for our method the coverage level is determined using the estimated uncertainty, while for the baseline it is based on a symmetric probability interval for case exclusion around 0.5. In other words, for varying $\delta$ we rejected cases with probability in $[0.5-\delta, 0.5+\delta]$.\label{fig:evoComparison}}
\end{figure}

Considering a standard classifier trained, e.g., by minimizing the binary cross entropy function as presented by~\cite{Guendel2018}, we emphasize that one cannot effectively use the probability measure alone to effectively perform sample rejection. We investigated this by eliminating samples with predicted probability close to a predefined decision boundary $d$. More details can be seen in Figure~\ref{fig:evoComparison} on the example of pleural effusion. 

\subsubsection{Relating Predictive Uncertainty with Reader Uncertainty}
We provide an analysis of the properties of the estimated predictive uncertainty on the example of pleural effusion based on the ChestX-Ray8 dataset. For the testing set of 689 cases $\mathcal{D}^{test}$ that was reread by a committee of 4 experts in consensus, we define the so called \emph{critical set/subset} $\mathcal{C} \subseteq \mathcal{D}^{test}$: containing only cases for which the original label established by~\cite{Wang2017} from the radiographic report via NLP was changed/flipped by the committee. According to the committee, this set contained both easier cases (for which the NLP technology presumably failed to extract the correct information from the radiographic report), and more difficult cases with subtle or atypical pleural effusions. 

\begin{figure}[t]
\centering
\includegraphics[width=10cm]{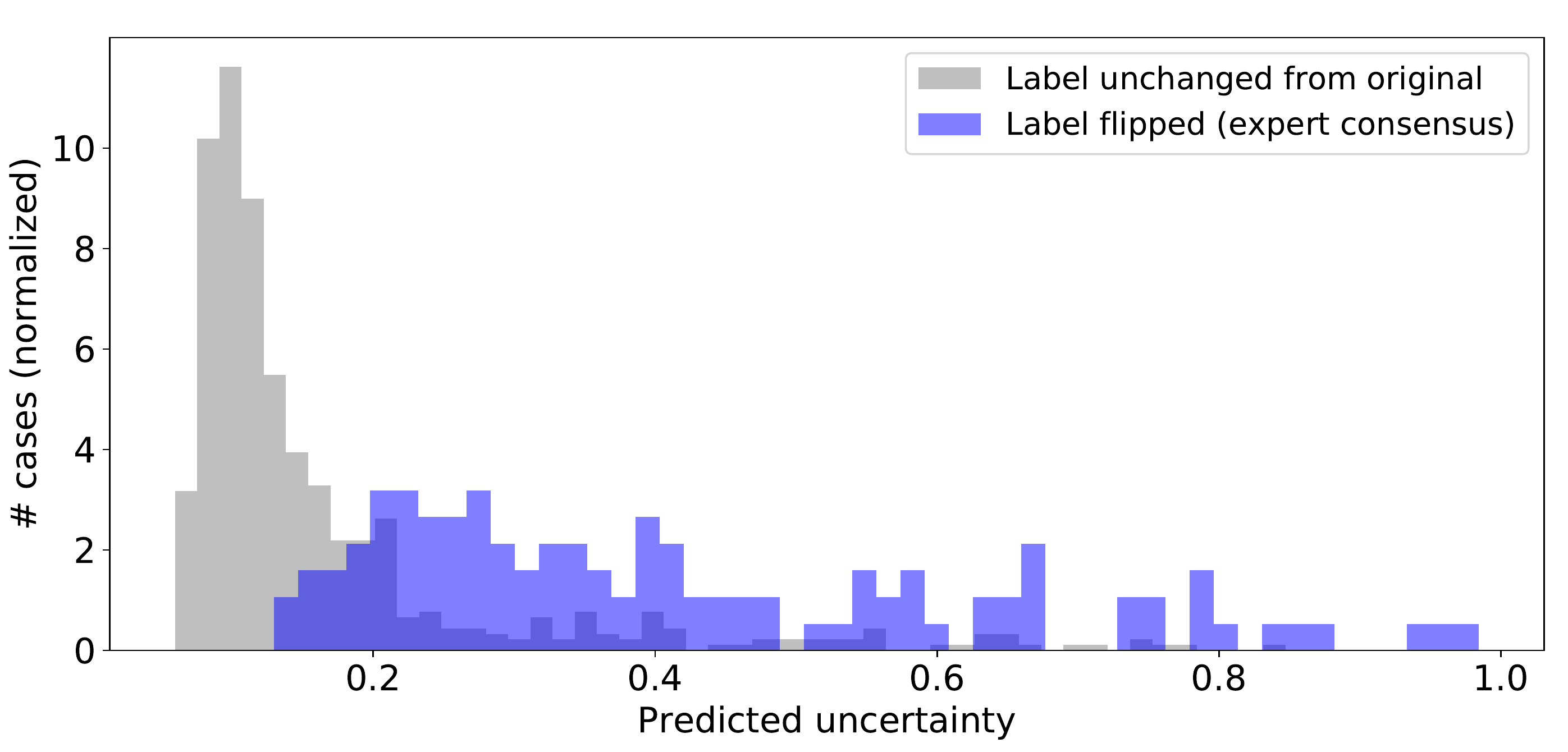}
\caption{\textbf{Left}: Distribution of the learning uncertainty measure for the 689 test images of ChestX-Ray8~\citep{Wang2017}. Cases of the critical set $\mathcal{C}$ are generally associated with a higher predictive uncertainty.\label{fig:consensus}}
\end{figure}

\begin{figure*}[t!]
\centering
\subfloat[$\hat{u},\hat{p}=0.90,0.45$]{
\includegraphics[height=5cm]{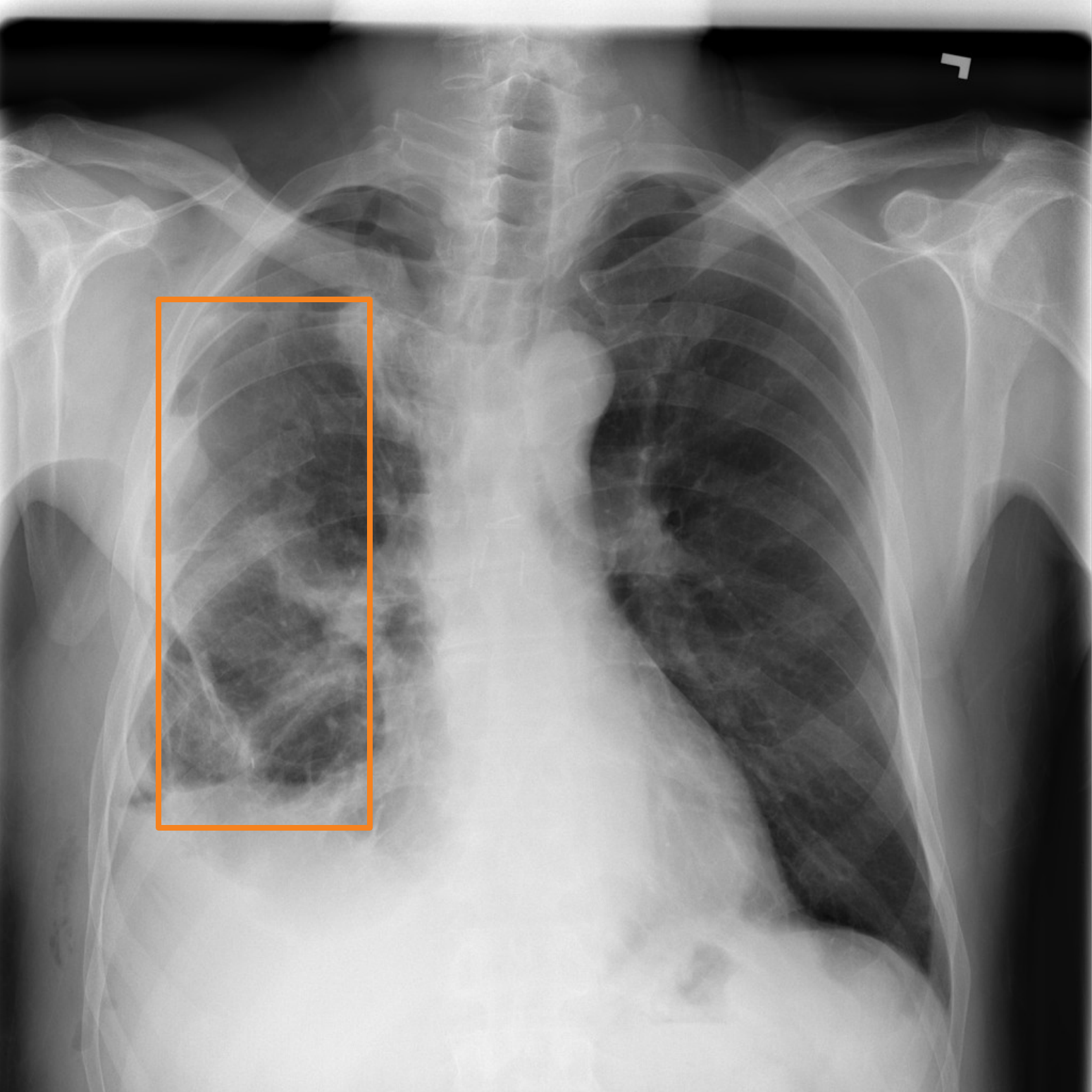}
\label{subfig:1}
}
\subfloat[$\hat{u},\hat{p}=0.93,0.48$]{
\includegraphics[height=5cm]{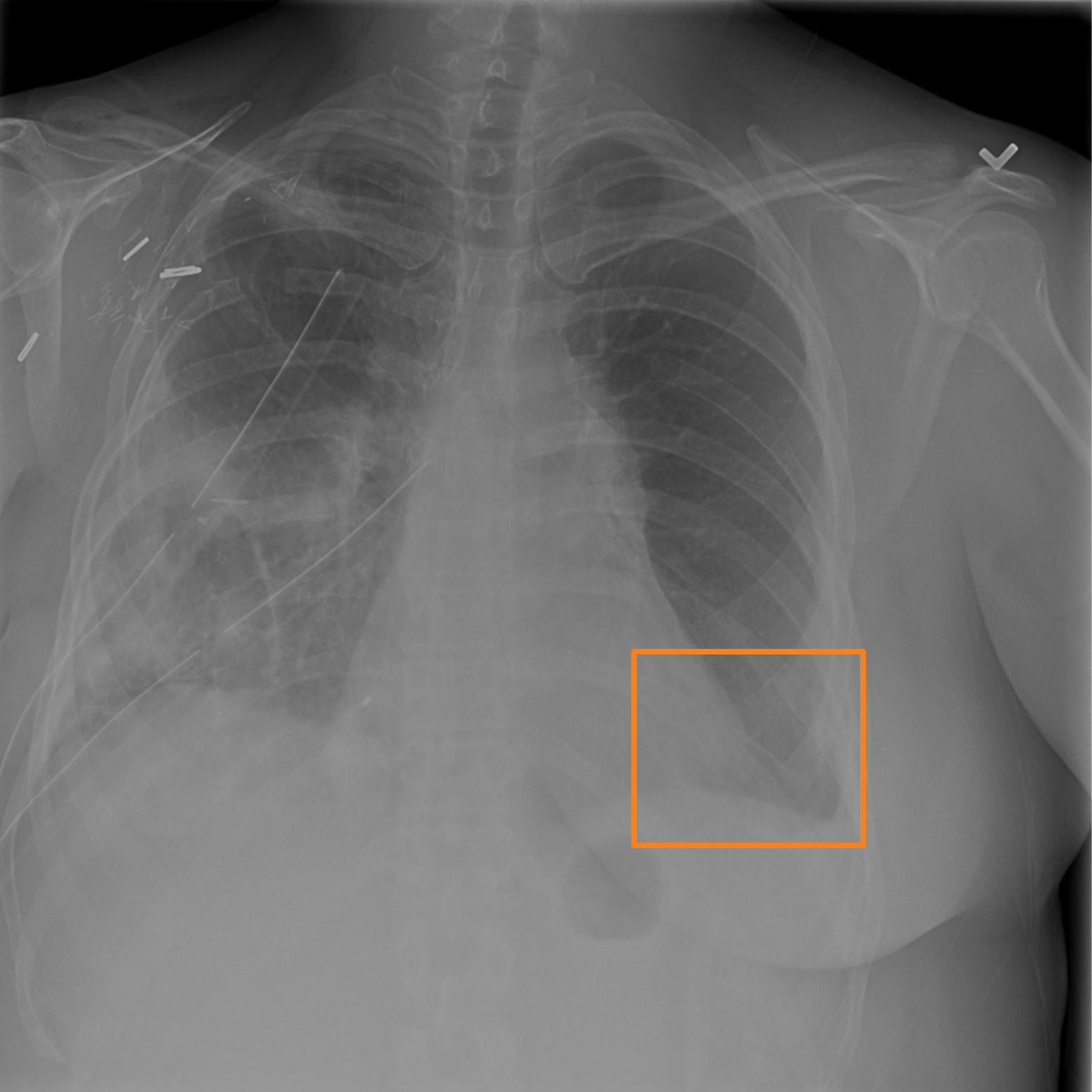}
\label{subfig:2}
}\\
\subfloat[$\hat{u},\hat{p}=0.54,0.65$]{
\includegraphics[height=5cm]{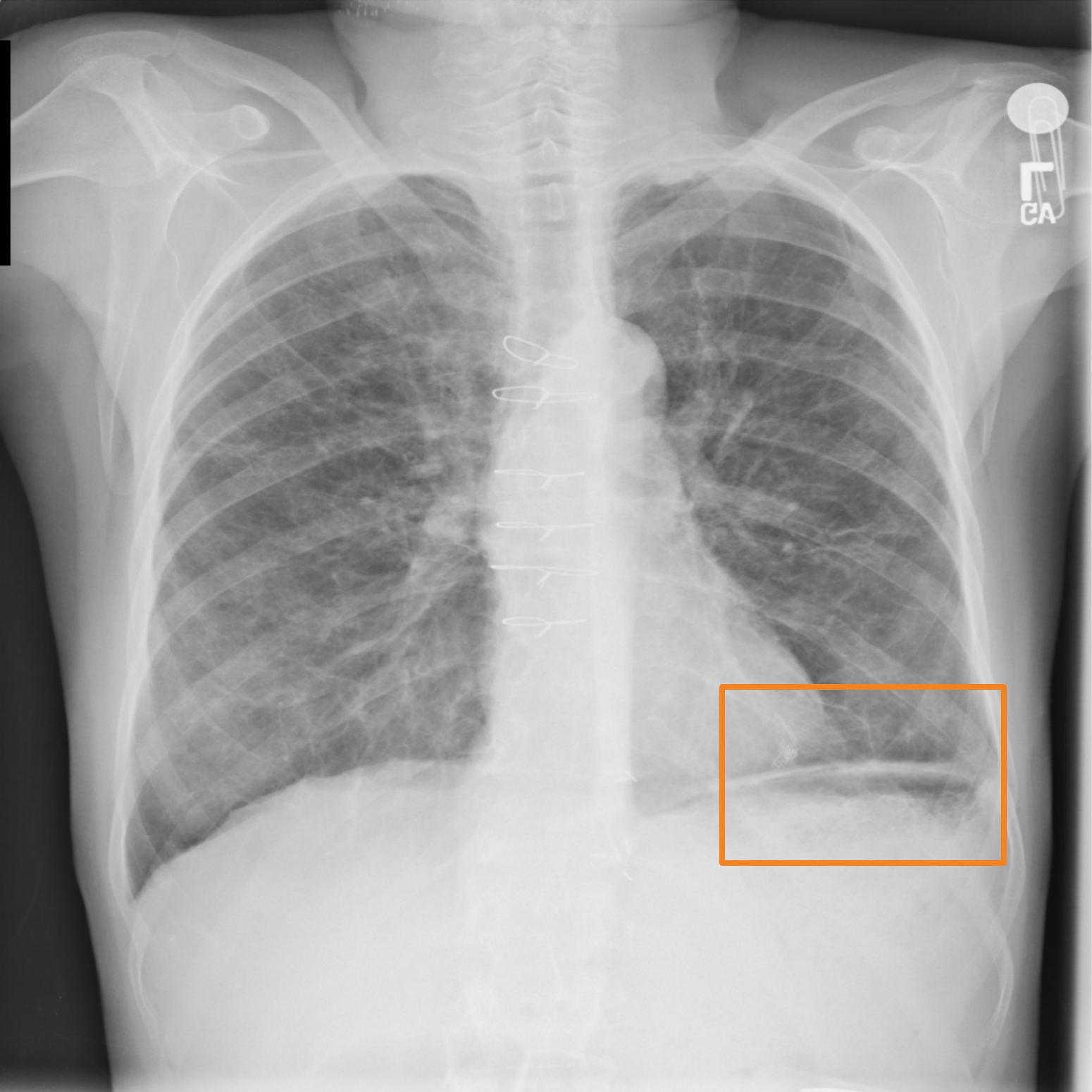}
\label{subfig:3}
}
\subfloat[$\hat{u},\hat{p}=0.11,0.05$]{
\includegraphics[height=5cm]{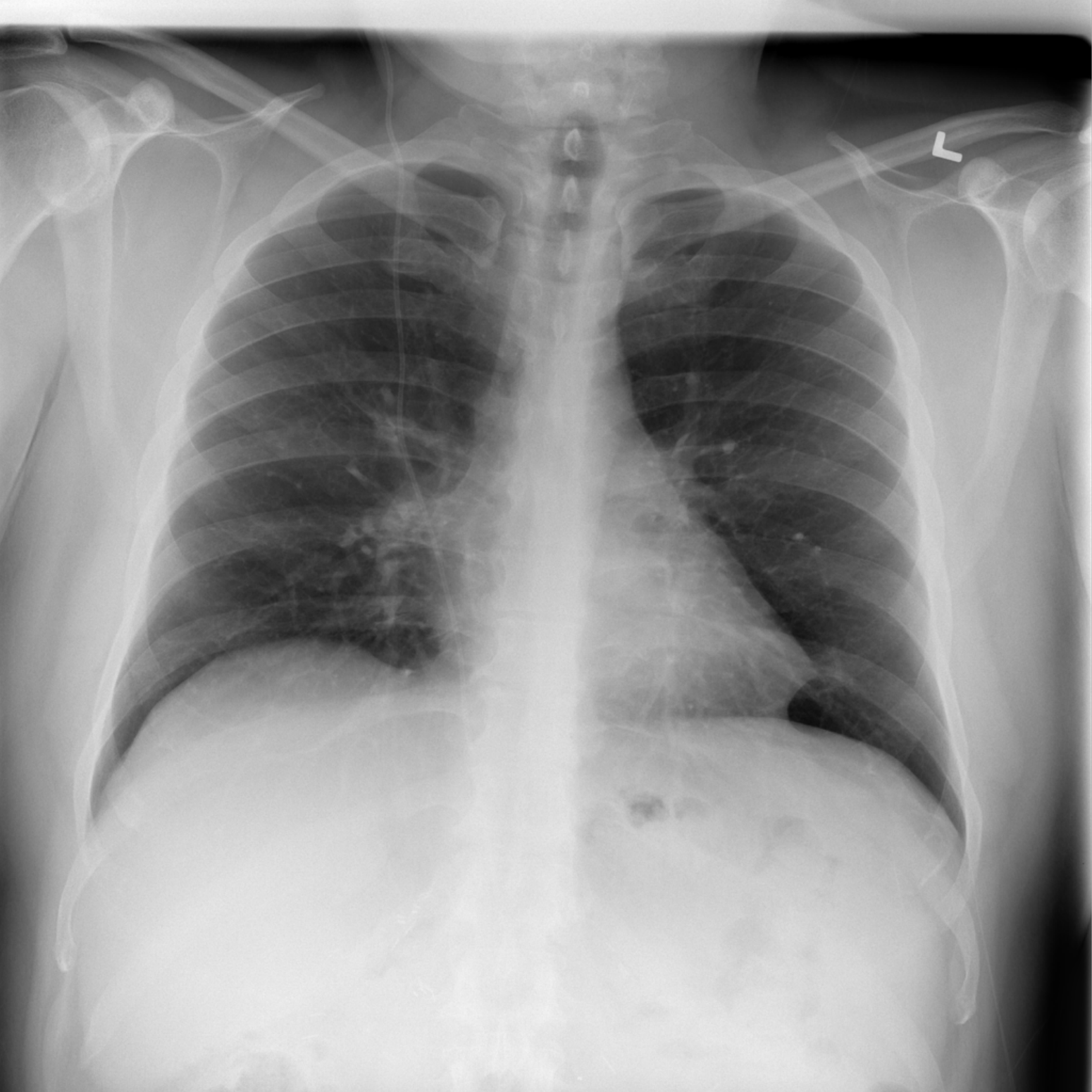}
\label{subfig:4}
}
\caption{Example images from the ChestX-Ray8 testing set on the example of pleural effusion ($\hat{u}$ denotes the estimated predictive uncertainty while $\hat{p}$ denotes the estimated probability on the presence of pleural effusion). The orange squares highlight the affected regions. Figures~\ref{subfig:1}, \ref{subfig:2} and \ref{subfig:3} show positive cases of the critical set $\mathcal{C}$ that have a higher predictive uncertainty. This may be explained by the atypical appearance of accumulated fluid in~\ref{subfig:1}, and the poor contrast of image~\ref{subfig:2}. Figure~\ref{subfig:4} shows a high confidence case with no pleural effusion.\label{fig:examples}}
\end{figure*}

In Figure~\ref{fig:consensus}, we empirically show that the predictive uncertainty measure correlates with the decision of the committee to change the image label. In other words, for cases in $\mathcal{D}^{test} \setminus \mathcal{C}$ (that preserve the original label) our algorithm yields lower uncertainty estimates (average 0.16) at an average AUC of 0.976 (coverage of 100\%). In contrast, on cases of the critical subset $\mathcal{C}$, the algorithm showed higher uncertainties distributed between 0.1 and the maximum value of 1 (average 0.41). This indicates the ability of the algorithm to recognize the cases where annotation errors occurred in the first place (through NLP or human reader error). In Figure~\ref{fig:examples} we show several qualitative examples.\smallskip

\subsubsection{Uncertainty-driven Bootstrapping} 

Using uncertainty driven bootstrapping one can filter the training data, i.e., remove a fraction of training cases with highest uncertainty, with the goal to reduce label noise. On the example of pleural effusion, based on the ChestX-Ray8 dataset, we show the one can retrain the system on the remaining data and achieve better performance on an unseen dataset. Performance is reported as a triple of values [\textit{AUC}; \textit{F1-score} (for the positive class); \textit{F1-score} (for the negative class)]. 

After the initial training, the baseline performance of our method was measured at $[0.89; 0.60; 0.92]$ on the testing set (including all 689 cases). We then constructed different version of the training set $\mathcal{D}_{0.05}$, $\mathcal{D}_{0.10}$, $\mathcal{D}_{0.15}$ by eliminating $5\%$, $10\%$ and $15\%$ of the training data. The metrics on the testing set improved to $[0.90; 0.68; 0.92]$ when retraining on $\mathcal{D}_{0.05}$, $[0.91; 0.67; 0.94]$ when retraining on $\mathcal{D}_{0.10}$ and finally $[0.93; 0.69; 0.94]$, when retraining on $\mathcal{D}_{0.15}$. This significant increase demonstrates the potential of this approach to improve the robustness of the model to label noise.

\subsection{View-classification on Abdominal Ultrasound Images}

A standard abdominal US examination typically consists of ten standard view classifications and their corresponding measurements from five structures of abdomen (right hepatic lobe, left hepatic lobe, right kidney, left kidney, and spleen) at two orientations (longitudinal/transverse). Here we focus on the longitudinal view-classification of the left versus the right kidney. 

\subsubsection{Dataset and Setup}
To demonstrate the efficacy of our proposed approach in leveraging the predictive uncertainty, we trained a state-of-the-art binary classifier to be used as baseline. For this we selected the DenseNet121 architecture~\citep{Huang2017}. For training and validation of the classification framework, we used $20,556$ kidney US images acquired retrospectively from 706 subjects. For testing, $3390$ US images ($108$ subjects) were used. The images were acquired both as longitudinal sequence and as single frame. The view information was assigned manually by at least one expert urologist either at the time of acquisition or retrospectively. As a pre-processing step for the framework, the images were resampled to $0.5$~mm resolution and resized to $256\times256$ pixels. Finally, a mask was applied to each image to hide any text or icon-based information that could indicate the organ in the view. 

The baseline performance of the classifier was measured at a ROC-AUC of 0.974. This is a competitive value, comparable to previous state-of-the-art results reported in \citep{xu2018less}, where a simultaneous view-classification and measurement framework for abdominal US exams was presented. To train the model for estimating predictive uncertainty we used the same architecture as for the baseline classifier and similar training meta-parameters as for the chest radiograph experiments.

\subsubsection{Uncertainty-driven Sample Rejection}
\label{subsubsec:samplerejection}
\begin{figure}[t]
\centering
\includegraphics[width=10cm]{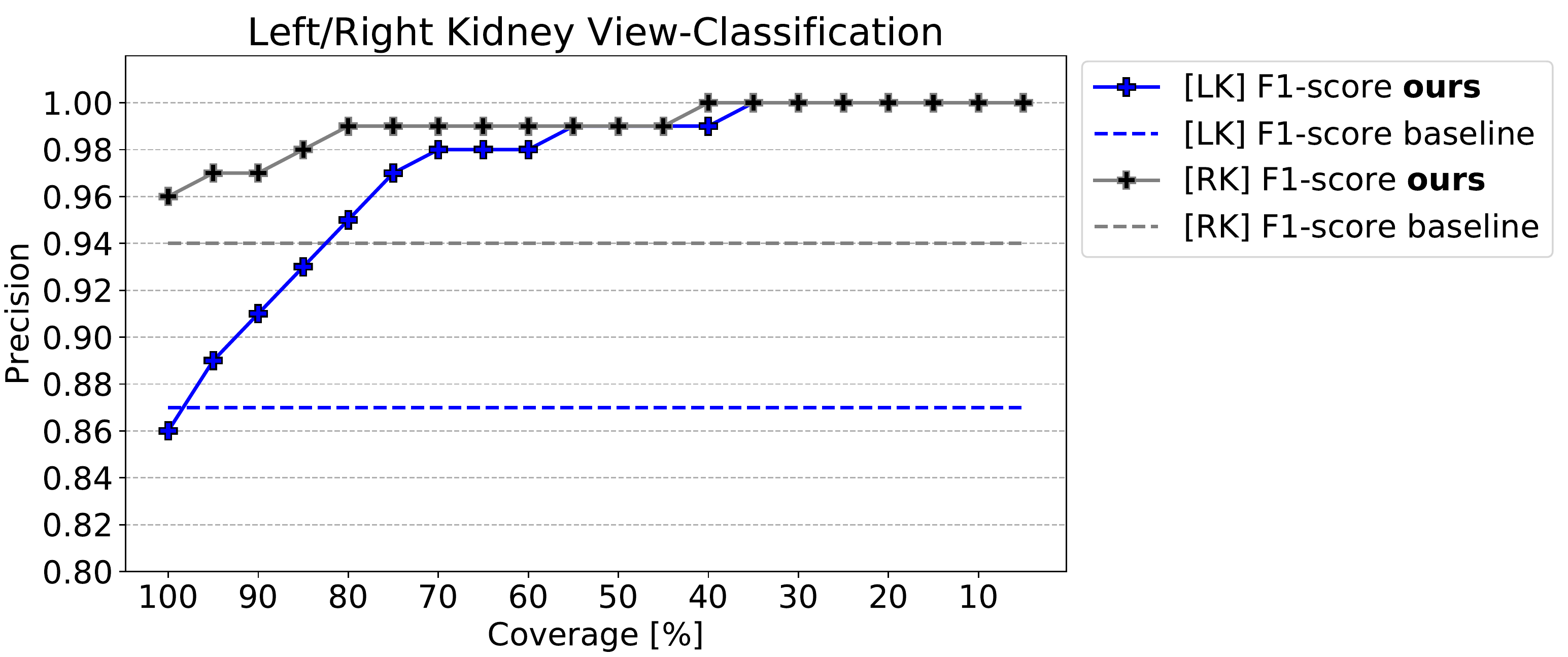}
\caption{Precision for the left kidney (LK) and right kidney (RK) classes as a function of the sample coverage, on the abdominal US view-classification example. The baseline performance is determined using a state-of-the-art deep convolutional classifier~\citep{Huang2017}. The working point of the reference method is chosen such that the average of the per-class precision scores is maximized. For our model, the decision threshold was set at 0.5.\label{fig:ussamplereject}}
\end{figure}

\begin{figure*}[t!]
\centering
\subfloat[$\hat{u},\hat{p}=0.11, 0.05$ (LK)]{
\includegraphics[width=5.5cm]{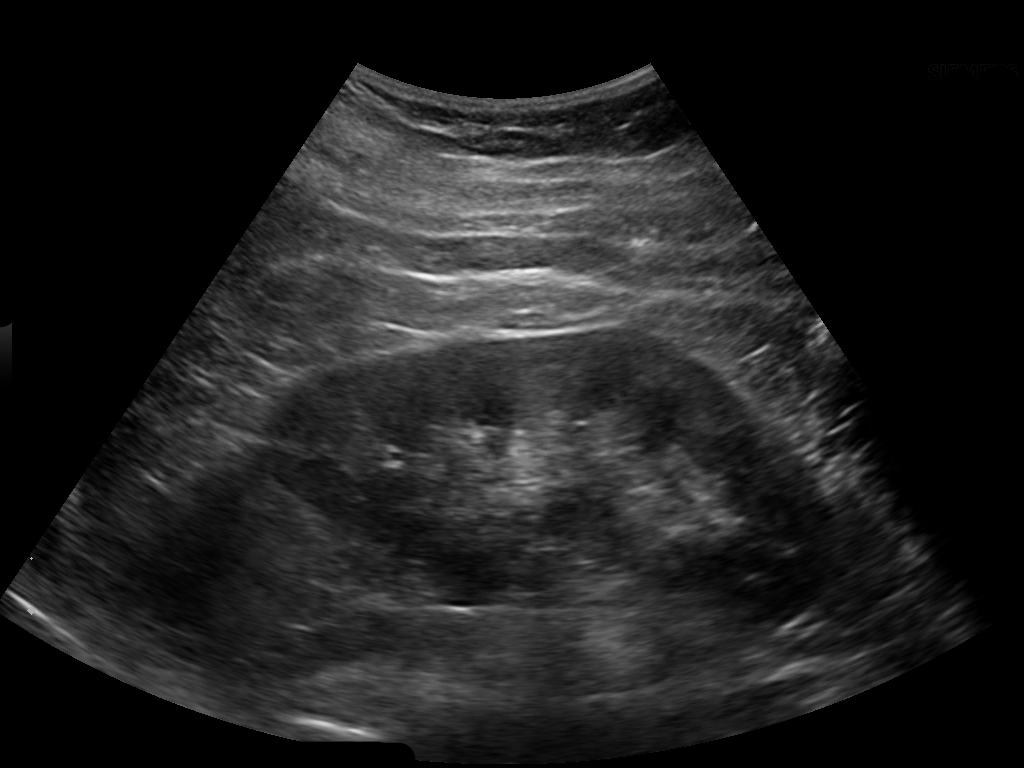}
\label{usubfig:1}
}
\subfloat[$\hat{u},\hat{p}=0.09, 0.96$ (RK)]{
\includegraphics[width=5.5cm]{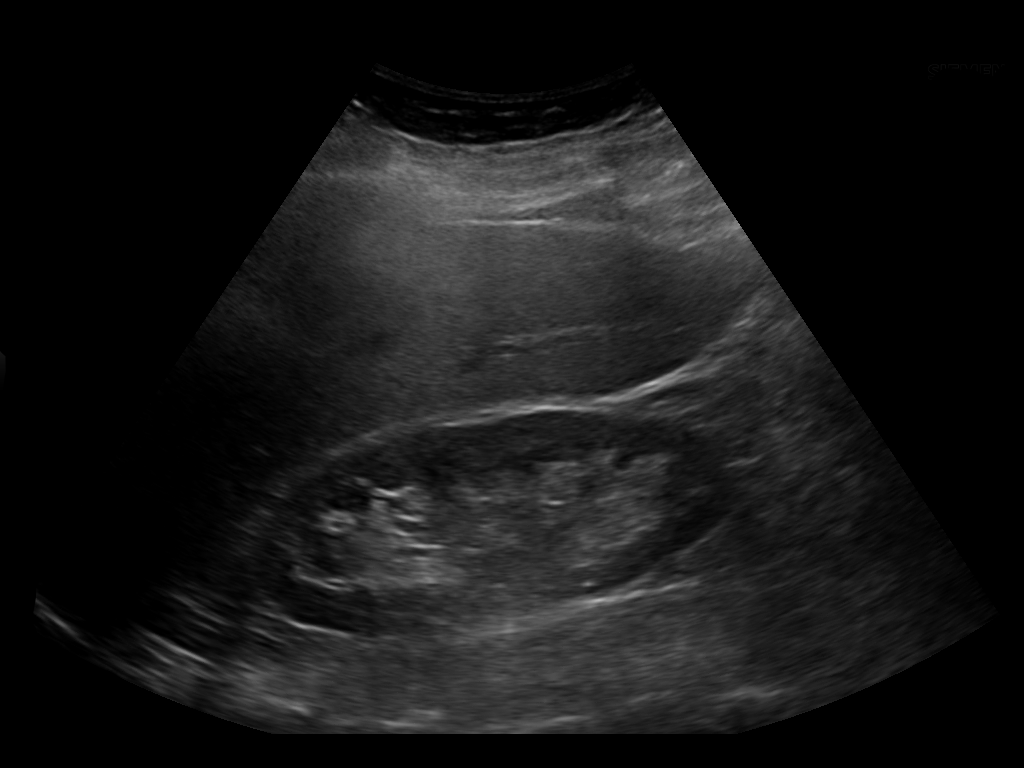}
\label{usubfig:2}
}\\
\subfloat[$\hat{u},\hat{p}=0.50, 0.40$ (LK)]{
\includegraphics[width=5.5cm]{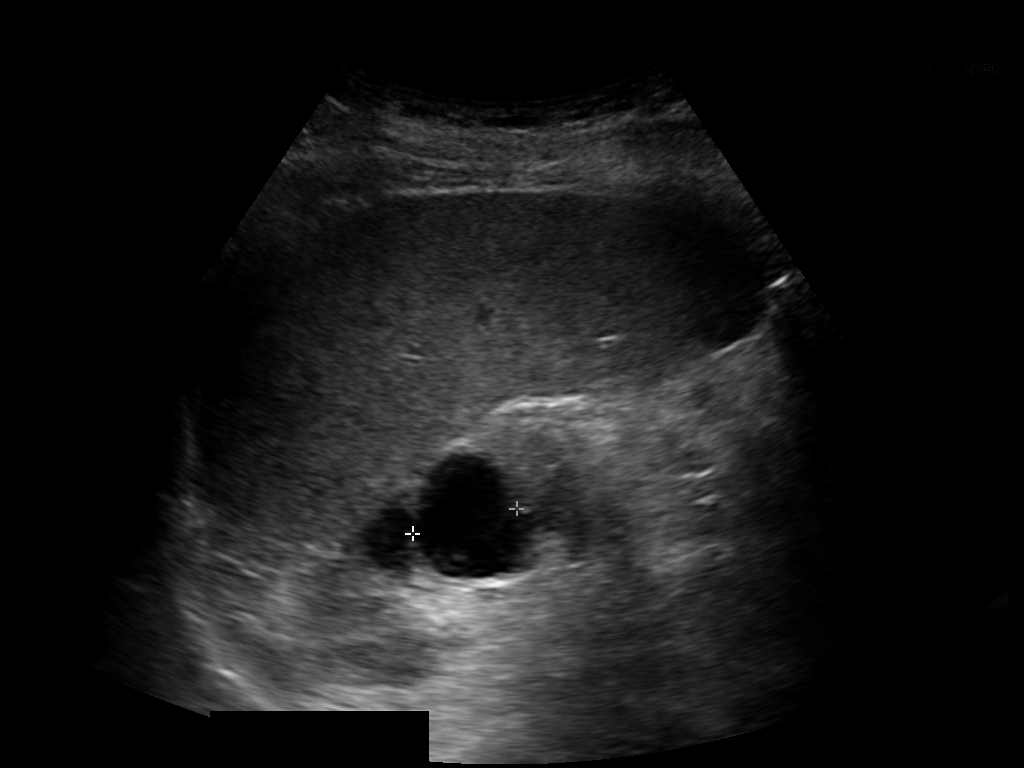}
\label{usubfig:3}
}
\subfloat[$\hat{u},\hat{p}=0.47, 0.50$ (RK)]{
\includegraphics[width=5.5cm]{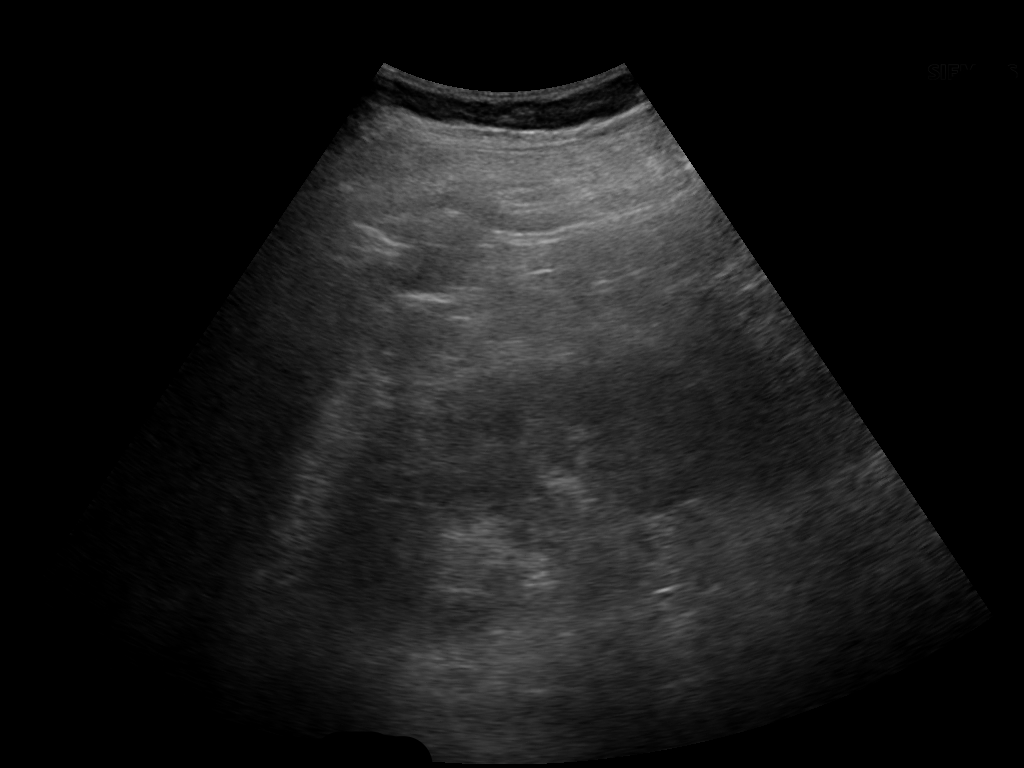}
\label{usubfig:4}
}
\caption{Example ultrasound images from the testing set ($\hat{u}$ denotes the estimated predictive uncertainty while $\hat{p}$ denotes the estimated probability of the image displaying the right kidney). On the left are shown images which according to the expert annotation display the left kidney (LK), while on the right we show images of the right kidney (RK). Figures~\ref{usubfig:1}, \ref{usubfig:2} display low uncertainty cases, while in Figures~\ref{usubfig:3}, \ref{usubfig:4} it is very challenging to distinguish the views (in Figure~\ref{usubfig:3} possibly due to the non-inclusion of the inferior aspect of the left kidney) - this is also reflected in a considerably higher predictive uncertainty value from the system.\label{fig:ultrasound}}
\end{figure*}

Following a similar strategy for rejecting samples with highest predictive uncertainty, one can significantly improve the classification performance of the trained model on the remaining cases, e.g., from a ROC-AUC of 0.972 at a coverage of 100\% to a ROC-AUC of 0.991 at a coverage of 80\% -- that is more than 10\% improvement in terms of precision. More details are shown in Figure~\ref{fig:ussamplereject}. Several example images with both high and low predictive uncertainty are shown in Figure~\ref{fig:ultrasound}.

\subsection{Brain metastases detection on MPRAGE images}
Automated detection and segmentation of small metastases in 3D MRI scans could support therapy workflows. However, this task remains challenging due, in part, to the imbalance between metastatic tissue and normal tissue in an MRI volume. Reliable detection or exclusion of metastases on 2D slices within a volumetric image processing pipeline can mitigate this imbalance. Thus, in this work, we focus on classifying 2D slices with metastases in MPRAGE volumes. 

\subsubsection{Dataset and Results}
We utilized a 2.5D encoder-decoder network to first obtain a segmentation mask showing potential areas of suspected metastases. The segmentation mask is subsequently used along with input slices as an input to a DenseNet121 model~\citep{Huang2017} to perform a slice-wise classification.

Our dataset included 480 contrast-enhanced MPRAGE image volumes from 442 patients treated primarily with stereotactic radiosurgery to one or more brain metastases. Metastasis gross tumor volumes, manually delineated in the course of standard clinical treatment, were reviewed for inclusion in the study. We excluded 47 cases where the planned treatment did not include all identifiable metastases. Further 13 cases were excluded due to non-standard orientations, field-of-views or imaging artifacts. The dataset was split into a training set (341 cases), a validation set (36 cases) and a test set (43 cases). In order to evaluate the performance of detecting small metastases, we selected all 16 patients from the test set with all metastatic lesions under 1 cm$^3$ in volume. The total number of annotated small metastases in this subset was 35.

\begin{figure}[t]
\centering
\includegraphics[width=11cm, height=6.0cm]{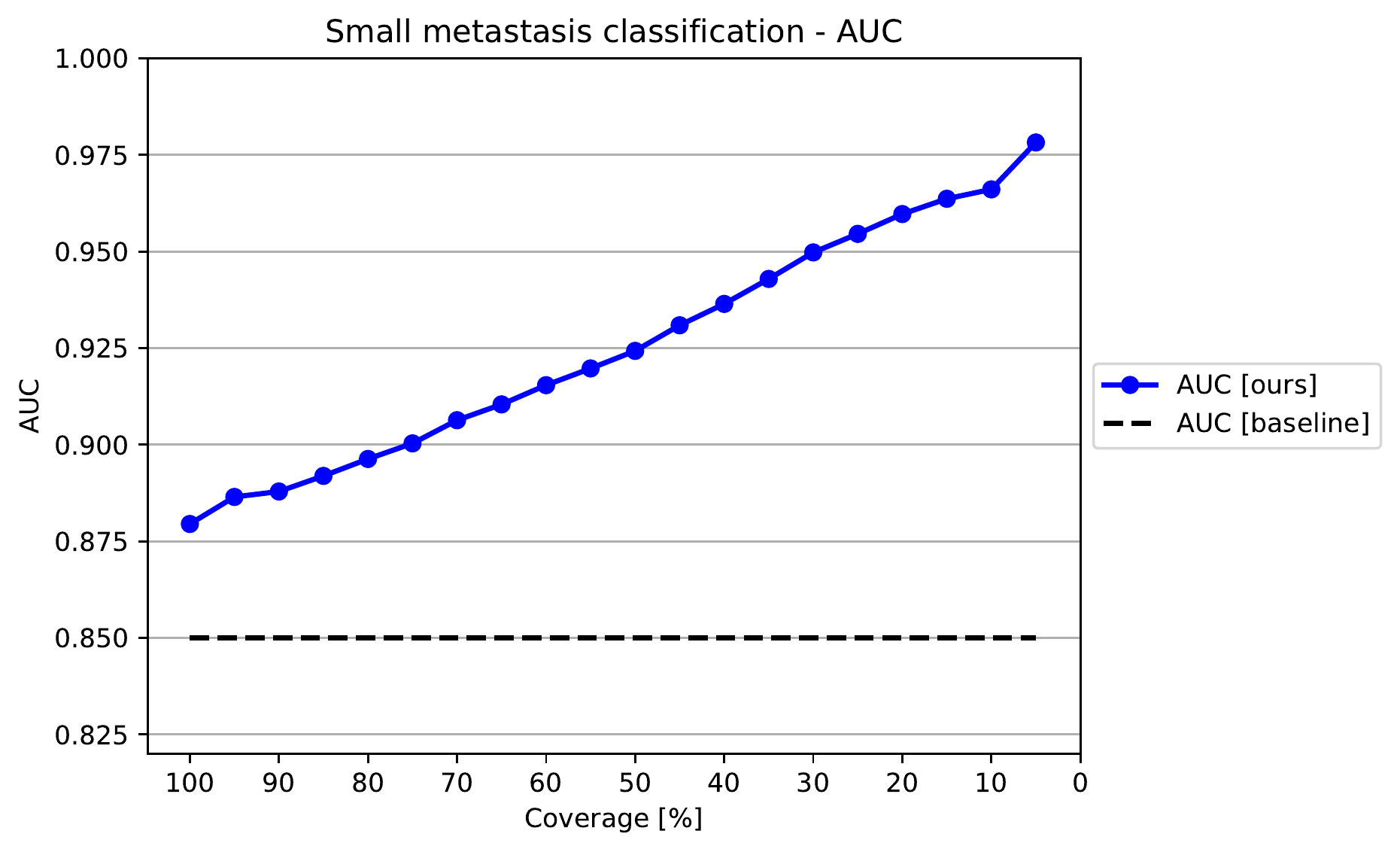}
\caption{ROC-AUC as a function of the sample coverage on the small brain metastases detection example. The baseline performance was established using state-of-the-art 2.5D deep convolutional neural network classifier.\label{fig:mrsamplereject}}
\end{figure}

\begin{figure*}[t!]
\centering
\subfloat[$\hat{u},\hat{p}=0.08, 0.96$]{
\includegraphics[width=5.5cm,height=6.5cm]{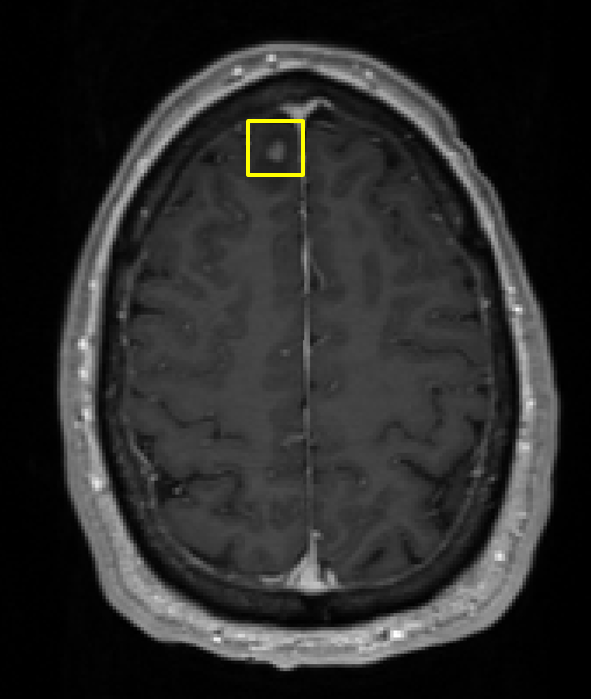}
\label{msubfig:1}
}
\subfloat[$\hat{u},\hat{p}=0.71, 0.35$]{
\includegraphics[width=5.5cm,height=6.5cm]{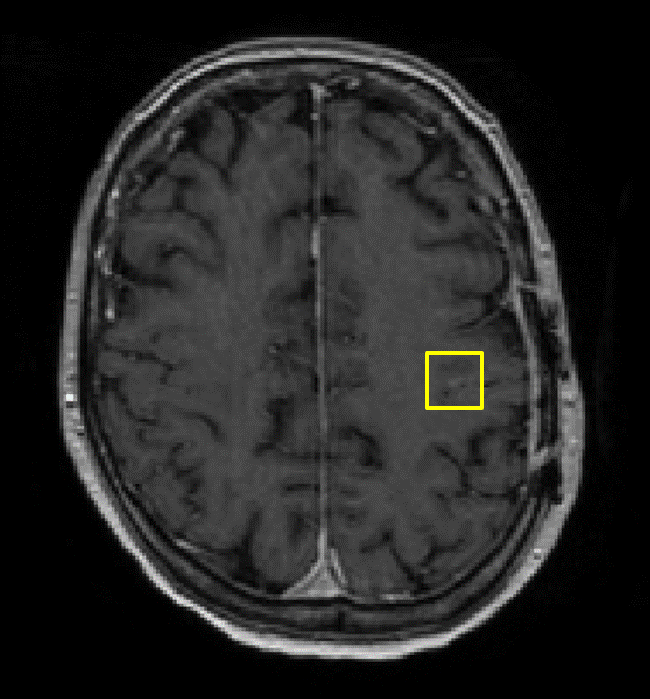}
\label{msubfig:2}
}
\caption{Example MPRAGE images from the test set ($\hat{u}$ denotes the estimated predictive uncertainty while $\hat{p}$ denotes the estimated probability of metastasis presence in the input image). Both images are metastases positive examples. The left image \ref{msubfig:1} shows a metastatic image in which the uncertainty measure is low. The right image \ref{msubfig:2} shows a metastatic image in which the uncertainty measure is high. The high uncertainty is likely due to ambiguity in visual representation of the metastasis. \label{fig:mrexample}}
\end{figure*}

To evaluate the efficacy of uncertainty-driven sampling rejection with high predictive uncertainty, we measured the classification performance of the trained model in different coverage settings. The baseline performance of the trained model without the proposed approach was 0.85 in ROC-AUC. Using our method ROC-AUC was increased to 0.88 at a coverage rate of 100\%, where no test samples were rejected by uncertainty-driven sampling described in Section~\ref{subsubsec:samplerejection}. This indicates the improved ability of our model to capture the underlying noise in the training labels. The classification performance was further increased up to a ROC-AUC of 0.925 at a coverage of 50\% and up to a ROC-AUC of 0.96 at a coverage of 20\%. More details are shown in Figure~\ref{fig:mrsamplereject}. Several example images with both high and low predictive uncertainty are shown in Figure~\ref{fig:mrexample}.

\section{Summary and Conclusion}

In conclusion, this paper presents an effective method for the joint estimation of class probabilities and predictive uncertainty. Extensive experiments were conducted on large datasets in the context chest of radiograph assessment, abdominal ultrasound view-classification and detection of small metastases in brain MR scans. We demonstrate that it is possible to achieve a significant increase in accuracy if sample rejection is performed based on the estimated uncertainty measure. For the assessment of chest radiographs, we highlight the capacity of the system to distinguish based on the uncertainty measure radiographs with correct and incorrect labels according to a multi-radiologist-consensus user study. Finally, we provide an insight into how to effectively use the predictive uncertainty to stratify the training dataset via bootstrapping to achieve higher accuracy on unseen data.\medskip

\subsection{Discussion and Directions of Future Research}

Based on these results and the potential impact of predictive uncertainty on system users, we believe that more research is required to address several open problems, including:
\begin{itemize}
    \item Investigation of additional sampling strategies which may allow to better capture the underlying distribution of the data and enable an improved estimate of predictive uncertainty
    \item Establishing a formal measure to quantify the quality of the estimated predictive uncertainty, distinguishing between reducible epistemic uncertainty and aleatoric data uncertainty. One step may be to relate the uncertainty to a consensus or majority decision of experts. In this paper, we made a first step in this direction for chest radiograph assessment based on a reader-study involving 4 board-certified radiologists.
    \item Based on the previous point, more research is needed into verifying how predictive uncertainty contributes to building user trust and reducing negative user bias. This is a key aspect to ensure that such systems are accepted and successfully used in daily practice.
\end{itemize}

\textbf{Acknowledgement} The authors thank the National Cancer Institute for access to NCI's data collected by the Prostate, Lung, Colorectal and Ovarian (PLCO) Cancer Screening Trial. The statements contained herein are solely those of the authors and do not represent or imply concurrence or endorsement by NCI.\medskip

\ifx\anonymize\undefined
\textbf{Disclaimer} The concepts and information presented in this paper are based on research results that are not commercially available.
\fi

\bibliography{paper.bib}

\end{document}